\documentclass[9pt,journal]{IEEEtran}
\usepackage{cite}
\usepackage{amsmath,amssymb,amsfonts}
\usepackage{graphicx}
\usepackage{textcomp}
\usepackage{algorithm, algorithmic}
\usepackage{colortbl}
\usepackage{epsfig}
\usepackage{epstopdf}
\usepackage{enumerate}
\usepackage{diagbox}

\renewcommand{\t}{{\mathsf{T}}}
\newtheorem{assume}{Assumption}

\newtheorem{Remark}{Remark}

\newtheorem{Problem}{Problem}
\newenvironment{Proof}{\noindent{\em Proof:\/}}{\hfill $\Box$\par}
\newtheorem{Theorem}{Theorem}
\newtheorem{Lemma}{Lemma}

\newtheorem{Claim}{Claim}

\newenvironment{ProofL2}{\noindent{\em Proof of Claim 1:\/}}{\hfill $\Box$\par}

\newcommand{\G}{\mathcal{G}}
\newcommand{\BG}{\mathcal{\bar G}}

\newcommand{\BV}{\mathcal{\bar V}}

\newcommand{\BE}{\mathcal{\bar E}}

\newcommand{\R}{\mathbb{R}}

\renewcommand{\t}{{\mathsf{T}}}

\def\BibTeX{{\rm B\kern-.05em{\sc i\kern-.025em b}\kern-.08em
		T\kern-.1667em\lower.7ex\hbox{E}\kern-.125emX}}

\begin{document}
\title{ Cooperative Global $\mathcal{K}$-exponential Tracking Control of Multiple Mobile Robots---Extended Version}
\author{Liang Xu, Youfeng Su, and He Cai  
		
\thanks{This work was supported by National Natural Science Foundation of China under Grant Nos. 62173092 and 62173149. \textit{(Corresponding author: Youfeng Su.)} }
\thanks{L. Xu is with College of Computer and Data Science, Fuzhou University, Fuzhou 350116, China, and also with Shenzhen Key Laboratory of Control Theory and Intelligent Systems, Southern University of Science and Technology, Shenzhen 518055, China (e-mail: liangxu0910@163.com).}
\thanks{ Y. Su is with College of Computer and Data Science, Fuzhou University, Fuzhou 350116, China (e-mail: yfsu@fzu.edu.cn). }
\thanks{H. Cai is with School of Automation Science and Engineering, South China University of Technology, Guangzhou 510641, China (e-mail: caihe@scut.edu.cn).}}
	
\maketitle

\begin{abstract}
	This paper studies the cooperative tracking control problem for multiple mobile robots over a directed communication network.
	First, it is shown that the closed-loop system is uniformly globally asymptotically stable under the proposed distributed continuous feedback control law, where an explicit strict Lyapunov function is constructed.
	Then, by investigating the convergence rate, it is further proven that the closed-loop system is globally $\mathcal{K}$-exponentially stable. Moreover, to make the proposed control law more practical, the distributed continuous feedback control law is generalized to a distributed sampled-data feedback control law using the emulation approach, based on the strong integral input-to-state stable Lyapunov function. Numerical simulations are presented to validate the effectiveness of the proposed control methods.
\end{abstract}



\section{Introduction}\label{sec-01}
Cooperative control has attracted tremendous attention for the past two decades as multi-agent systems exhibit many advantages over a single system, such as cost reduction, fault tolerance, working efficiency improvement, and so on \cite{book-2008-RW}. Among the extensive results regarding multi-agent systems, cooperative control of nonholonomic systems has been widely studied since many practical systems are subject to nonholonomic constraints, for instance, mobile robots \cite{book-2009-QZH}. 

Crucial works on cooperative tracking control of multiple nonholonomic mobile robots with a dynamic leader can be found in \cite{TAC-2008-DWJ,TR-2012-DWJ,IJSS-2015-PZX, TASE-2018-MZQ, Auto-2013-LTF, Auto-2014-LTF, TAC-2018-Loria, TAC-2020-Loria, ND-2017-TGP,TAC-2018-NB, IJRNC-2021-HGeng,CSL-2022-WNA,TIE-2015-YL,TAC-2023-YS}.
In \cite{TAC-2008-DWJ}, under the condition that the leader's angular velocity is persistently exciting (PE), a distributed controller was proposed to achieve cooperative tracking over a balanced communication network. Toward this direction, the same PE condition as in \cite{TAC-2008-DWJ} were considered in \cite{TR-2012-DWJ,IJSS-2015-PZX,CSL-2022-WNA,TAC-2023-YS}.
Specifically, several discontinuous leader-following distributed controllers were developed in \cite{IJSS-2015-PZX,TR-2012-DWJ} to deal with nonholonomic constraints for undirected communication networks. A continuous leader-following distributed controller using polar coordinates was proposed in \cite{CSL-2022-WNA} provided that the network graph is acyclic. The decentralized circular formation control was developed in \cite{TAC-2023-YS}. A two-stage nonlinear distributed control synthesis was proposed in \cite{Auto-2013-LTF,Auto-2014-LTF}  so that the follower's linear velocity would always be nonzero during transformation. In particular, the distributed tracking problem without global position measurement was  
investigated in \cite{Auto-2013-LTF}, which was further extended to time-varying communication networks in \cite{Auto-2014-LTF} by using the trajectory-based small gain technique. Papers \cite{TAC-2018-Loria,TAC-2020-Loria} loosened the PE assumption to the case that either the linear or the angular reference velocity is PE and the case that both the linear and the angular
reference velocity are zero, respectively, using an agent-to-agent design subject to a tree graph.  Papers \cite{ND-2017-TGP,IJRNC-2021-HGeng} worked on the consensus tracking problem without
PE constraints on the leader, where a virtual leader using
the convex combination on nonlinear manifolds was proposed based on an acyclic graph. Some attempts using the distributed observer approach were reported in \cite{TAC-2018-NB} assuming that the leader's angular velocity is PE, and in \cite{TIE-2015-YL,TASE-2018-MZQ}   assuming that the leader's linear velocity is PE, where an observer network was established instead of (or in addition to) the standard sensor network.




Note that all the aforementioned results rely on continuous feedback control, which may not always be feasible due to the limitation of computation and communication resources. In addressing this issue, sampled-data control has been explored \cite{IJC-2019-PZX, Auto-2017-LZX, TAC-2023-ZPP}. In \cite{IJC-2019-PZX}, the formation tracking problem for nonholonomic mobile robots was solved by event-triggered and sampled-data methods under undirected communication networks. \cite{Auto-2017-LZX} proposed a distributed sampled-data control strategy which can achieve both orientation and speed tracking asymptotically. Most recently, both event-triggered and self-triggered methods were developed in \cite{TAC-2023-ZPP} to solve the tracking problem for a single leader-follower pair of mobile robots. \cite{SCIS-Xu-Su-Cai-2023} further investigated the global event-triggered tracking control also for the single leader-follower pair.

In this paper, we address the cooperative global tracking problem of multiple mobile robots with one robot being the leader. We first propose a distributed continuous feedback control law based on only the sensor networks  to solve the cooperative global tracking problem, then we show that the closed-loop system is globally $\mathcal{K}$-exponentially stable.
Furthermore, to make the proposed control law more practical, the distributed continuous feedback control law is generalized to a distributed sampled-data feedback control law by the emulation approach. The main contributions of this paper are threefold:

Firstly, both the proposed continuous and sampled-data feedback control laws can accommodate general directed communication networks, which has relaxed the assumptions in many existing results, such as undirected networks \cite{TR-2012-DWJ,IJSS-2015-PZX},   balanced networks \cite{TAC-2008-DWJ}, and pure tree networks \cite{TAC-2018-Loria,TAC-2020-Loria}, acyclic networks \cite{CSL-2022-WNA,ND-2017-TGP, IJRNC-2021-HGeng}. 

Secondly, an explicit strict Lyapunov function is constructed for the closed-loop system under the distributed continuous feedback control law. Based on a joint analysis by several nonlinear control techniques, namely, the Lyapunov function strictification \cite{book-2009-Mazenc} and refinement \cite{Auto-2017-Maghenem} techniques for linear time-varying systems as well as the changing supply function technique for integral input-to-state stable (iISS)-Lyapunov function \cite{TAC-2010-Ito}, we are able to finely characterize the convergence rate of the tracking errors and prove global $\mathcal{K}$-exponential stability of the closed-loop system.

Thirdly, a novel sampled-data feedback control framework is given based on the strong iISS-Lyapunov function for the cascaded closed-loop system. The fact that strong iISS implies small input-to-state stability enables us to characterize the admissible sampling period by ensuring Lyapunov-like function decreased from one sampling instance to the next, for any sufficiently large time. Different from some previous results on cooperative sampled-data feedback control, say \cite{IJC-2019-PZX}, the communication network in this paper is allowed to be directed. In contrast to \cite{Auto-2017-LZX}, besides orientation and speed tracking, position tracking has further been achieved here.

The rest of this paper is organized as follows: Section \ref{sec-02} gives the problem formulation and some preliminaries. Section \ref{sec-Mresults} presents the distributed continuous and sampled-data feedback control laws, with an emphasis on the exponential convergence rate of the tracking errors. Simulation results and conclusions are provided in Sections \ref{sec-example} and \ref{sec-conclusion}, respectively.


\textit{Notations}: $\R, \R^{n}, \R^{m\times n}$ are the real number field, the $n$-dimensional Euclidean space, and the $m\times n$ real matrix space, respectively. $|x|$ denotes the absolute value of a real number $x$. $\Vert x\Vert$ denotes the Euclidean norm of a vector $x$. $\Vert A\Vert$ denotes the induced norm of a matrix $A$ by the Euclidean norm. For a vector $\xi=[\xi_{1},\dots,\xi_{n}]^{\t}\in\R^{n}$, $\mbox{diag}(\xi)$ represents $\mbox{diag}(\xi_{1},\dots,\xi_{n})$. $\lambda_{\min}(A)$ and $\lambda_{\max}(A)$ represent the minimum and maximum eigenvalues of a symmetric matrix $A$. A matrix $A\in\mathbb{R}^{n\times n}$ is called $\mathcal{M}$-matrix if every off-diagonal entry of $A$ is non-negative and all of its eigenvalues have positive real parts. 
A continuous function $\alpha:[0,\infty)\rightarrow[0,\infty)$ is of class $\mathcal{K}$, if it is strictly increasing and $\alpha(0)=0$. For any rational number $a$, $\lfloor a\rfloor$ denotes the maximum integer not greater than $a$.

\section{Problem formulation and Preliminaries}\label{sec-02}

Consider a group of $N$ nonholonomic mobile robots moving in a common $XY$-plane. For $i=1,\dots, N$, each robot is modeled as: 
\begin{align}\label{mb-01}
	\dot x_{i}=v_{i}\cos(\theta_{i}),~\dot y_{i}=v_{i}\sin(\theta_{i}),~\dot\theta_{i}=\omega_{i} 
\end{align}
where $[x_{i}, y_{i}]^{\t}\in\R^{2}$ denote the Cartesian coordinates (position) of the center of $i$-th robot, $\theta_{i}\in\R$ represents the $i$-th robot heading angle (orientation) with respect to $X$-axis, $v_{i},\omega_{i}\in\R$ are the linear and angular velocities of $i$-th robot, respectively. We regard $v_{i}$ and $\omega_{i}$ as the control inputs of $i$-th robot. For these $N$ robots, the desired reference trajectory is generated by the following system:
\begin{align}\label{re-01}
	\dot x_{0}=v_{0}\cos(\theta_{0}),~\dot y_{0}=v_{0}\sin(\theta_{0}),~\dot \theta_{0}=\omega_{0}
\end{align}
where $[x_{0}, y_{0}]^{\t}\in\R^{2}$ and $\theta_{0}\in\R$ are desired reference position and orientation, respectively. Here $v_{0}(t)$ and $\omega_{0}(t)$ denote the inputs of  system \eqref{re-01}, which are some known time-varying functions, and for simplicity, can be accessed by all $N$ robots.

The system composed of \eqref{mb-01} and \eqref{re-01} can be regarded as a multi-agent system with \eqref{re-01} as the leader (labeled by 0) and the $N$ robots of \eqref{mb-01} as the followers, respectively. The communication network among all robots are described by a directed graph $\BG=(\BV, \BE)$, where $\BV=\{0,1,\dots,N\}$ and an edge $(i,j)\in\BE$ means that the $j$-th robot can receive the information from  the $i$-th robot. The control is to be distributed in the sense that the information of the position and the angle can only be transmitted through the network. Then, the leader-following cooperative tracking problem is described as follows:
\begin{Problem}\label{pro-01}
Given a communication topology graph $\BG$, consider systems \eqref{mb-01} and \eqref{re-01}. For each $i$-th robot, design an appropriate distributed control law $(v_i,w_i)$, $i=1,\dots,N$, such that, for any initial values $x_{i}(t_{0}), y_{i}(t_{0}), \theta_{i}(t_{0})$, the trajectories of systems \eqref{mb-01} and \eqref{re-01} exist for all $t\geq t_0$, and 
	\begin{align}\label{aim-00}
		\lim\limits_{t \to \infty}([x_{i}(t), y_{i}(t), \theta_{i}(t)]-[x_{0}(t),y_{0}(t), \theta_{0}(t)])=0.
	\end{align}
\end{Problem}

\begin{Remark}
	Some other cooperative control objectives can also be transformed into \eqref{aim-00}. For instance, \cite{TIE-2015-YL, TASE-2018-MZQ} aimed to design   distributed leader-following formation control laws such that $\lim\nolimits_{t \to \infty}([x_{i}(t), y_{i}(t), \theta_{i}(t)]-[x_{0}(t),y_{0}(t), \theta_{0}(t)])=[p^{x}_{i0}, p^{y}_{i0},0]$ for any given geometric pattern $\mathcal{P}=\{[p^{x}_{i0}, p^{y}_{i0}]:~i=1,\dots,N\}$, where $[p^{x}_{i0}, p^{y}_{i0}]^{\t}$ represents the desired relative position between the $i$-th robot and the leader. Note that such leader-following formation objective can be converted to \eqref{aim-00} by state transformation $x^{\prime}_{i}=x_{i}-p^{x}_{i0}$, $y^{\prime}_{i}=y_{i}-p^{y}_{i0}$, and $\theta^{\prime}_{i}=\theta_{i}$, with $[x^{\prime}_{i}, y^{\prime}_{i}, \theta^{\prime}_{i}]^{\t}$ being the new state. 
\end{Remark}
	
Some standard assumptions are given as follows.
\begin{assume}\label{ass-02}
The leader's states $x_{0}$ and $y_{0}$ are bounded, i.e., there exists a constant $M>0$, such that
$
\max_{t\geq 0}\{|x_{0}(t)|, |y_{0}(t)|\}\leq M
$.
\end{assume}
	
\begin{assume}\label{ass-03}
The leader's angular velocity $\omega_{0}(t)$ is differentiable and both itself and its derivative are bounded, i.e., there exists  $\bar\omega>0$ such that 
$
\max_{t\geq 0}\{|\omega_{0}(t)|, |\dot\omega_{0}(t)|\}\leq \bar\omega.
$
Moreover, there exist positive numbers $T$ and $\mu$, such that
$
\int_{t}^{t+T}\omega_{0}^{2}(\tau){\rm d}\tau\geq \mu, \ \forall t\geq 0.
$
\end{assume}
\begin{assume}\label{ass-04}
	The graph $\mathcal{\bar G}$ contains a direct spanning tree with node $0$ as its root.
\end{assume}
	
\begin{Remark}
Assumption \ref{ass-02} admits various robot motions within a bounded area (see \cite{TR-2012-DWJ}, \cite{TAC-2018-NB}, \cite{IJC-2019-PZX}), such as the circular motion (see \cite{TAC-2023-YS}). Assumption \ref{ass-03} indicates that the leader's angular velocity $\omega_{0}$ is PE, which has been widely adopted in tracking, consensus, and formation problem for mobile robots (see \cite{TAC-2008-DWJ,IJSS-2015-PZX,TR-2012-DWJ,TAC-2018-NB,CSL-2022-WNA,TAC-2023-YS, IJC-2019-PZX}). Unlike the leaderless consensus scenario \cite{CSL-2023-DLPSN} where the PE-type signals can be freely preassigned, it was common and important for the leader's velocity to satisfy some PE conditions for the leader-following tracking scenario. Assumption \ref{ass-04} is standard for multi-agent systems over static communication networks (see \cite{Auto-2013-LTF,TAC-2012-SYF}).
\end{Remark}

\begin{Remark}
	As in \cite{book-2008-RW, Auto-2013-LTF, TAC-2023-YS} and many other existing works, it is assumed that the leader's velocity is known by all the followers which is reasonable for most typical application scenarios.
	While, for the case that the leader's velocity is not accessible to all follower robots, one may employ the distributed observer approach \cite{TIE-2015-YL} or the discontinuous approach \cite{TR-2012-DWJ}.
\end{Remark}

Let us introduce the error models and convert Problem \ref{pro-01} via some coordinate transformations. Inspired by the transformation that converts multiple robot systems into multiple chained-form systems \cite{book-2009-QZH}, for each robot $i$, consider the rotation transformation:
\begin{align}\label{Tr-01}
	\begin{bmatrix} \tilde{x}_{i} \\ \tilde{y}_{i} \end{bmatrix}=
	\begin{bmatrix}
		\cos(\theta_{i}) & \sin(\theta_{i})\\
		-\sin(\theta_{i}) & \cos(\theta_{i})
	\end{bmatrix}
	\begin{bmatrix} x_{i} \\ y_{i} \end{bmatrix},~i=0,1,\dots,N.
\end{align}
Then, systems \eqref{mb-01} and \eqref{re-01} can be transformed into  
\begin{align}\label{mb-02}
	\dot\theta_{i}=\omega_{i},~\dot{\tilde x}_{i}=v_{i}+\omega_{i}\tilde y_{i},~\dot{\tilde y}_{i}=-\omega_{i}\tilde x_{i}, \ i=0,1, \dots, N.
\end{align}
Let
$
\bar \theta_{i}=\theta_{i}-\theta_{0},~\bar x_{i}=\tilde x_{i}-\tilde x_{0},~ \bar y_{i}=\tilde y_{i}-\tilde y_{0}
$,
which yields
\begin{align}\label{error-sys}
	\dot{\bar \theta}_{i}&=\omega_{i}-\omega_{0} \notag\\
	\dot {\bar x}_{i}
	&=(\omega_{i}-\omega_{0})\tilde y_{i}+\omega_{0}\bar y_{i}+v_{i}-v_{0} \notag\\
	\dot {\bar y}_{i}
	&=-(\omega_{i}-\omega_{0})\tilde x_{i}-\omega_{0}\bar x_{i}.
\end{align}
Then, we have the following results:
	
\begin{Lemma}\label{lemma-02}
If $\bar x_{i}$, $\bar y_{i}$, and $\bar\theta_{i}$ of \eqref{error-sys} satisfy
$
\lim_{t \to \infty}[\bar x_{i}(t), \bar y_{i}(t),$ $\bar \theta_{i}(t)]=0,~i=1,\dots, N,
$
then, \eqref{aim-00} holds.
\end{Lemma}
	
\begin{Proof}
Notice that $\lim_{t \to \infty} (\theta_{i}(t)-\theta_{0}(t))= \lim_{t \to \infty} \bar{\theta}_i(t)=0$. On the other hand, from \eqref{Tr-01}, with some calculations, it gives  
	\begin{align*}
		&	\begin{bmatrix} x_{i}- x_{0} \\ y_{i}- y_{0}\end{bmatrix}
		=\begin{bmatrix} \cos (\theta_{i}) & -\sin (\theta_{i}) \\ \sin (\theta_{i}) & \cos (\theta_{i}) \end{bmatrix}
		\begin{bmatrix} \bar{x}_{i}  \\ \bar{y}_{i}  \end{bmatrix}-\begin{bmatrix} \cos (\theta_{i}) & -\sin (\theta_{i}) \\ \sin (\theta_{i}) & \cos (\theta_{i}) \end{bmatrix}\notag\\
		&~~~\times\begin{bmatrix} \cos (\bar\theta_{i}+\theta_{0})-\cos (\theta_{0}) & \sin (\bar\theta_{i}+\theta_{0})-\sin (\theta_{0}) \\ -(\sin (\bar\theta_{i}+\theta_{0})-\sin (\theta_{0})) & \cos (\bar\theta_{i}+\theta_{0})-\cos (\theta_{0}) \end{bmatrix}\begin{bmatrix} x_{0} \\ y_{0} \end{bmatrix}.
	\end{align*}
	This further implies that $\lim\nolimits_{t \to \infty}([x_{i}(t), y_{i}(t)]-[x_{0}(t),y_{0}(t)])=0$ provided that $\lim\nolimits_{t \to \infty} [\bar x_{i}(t), \bar y_{i}(t), \bar{\theta}_i(t)] =0.$ 
\end{Proof}

By virtue of Lemma \ref{lemma-02}, to achieve \eqref{aim-00}, it suffices to find $(v_i,w_i)$ to stabilize system \eqref{error-sys}. In this sense, we say Problem \ref{pro-01} is converted to the stabilization problem of system \eqref{error-sys}. However, due to the communication constraints, the signals $\bar x_{i}$ and $\bar\theta_{i}$ have also to be accessed in distributed sense. So we further define the virtual errors 
$
	e_{\theta_{i}}=\sum_{j=0}^{N}a_{ij}(\theta_{i}-\theta_{j})=\sum_{j=1}^{N}a_{ij}(\bar\theta_{i}-\bar\theta_{j})+a_{i0}\bar\theta_{i}$ and $
	e_{\tilde x_{i}}=\sum_{j=0}^{N}a_{ij}(\tilde x_{i}-\tilde x_{j})=\sum_{j=1}^{N}a_{ij}(\bar x_{i}-\bar x_{j})+a_{i0} \bar x_{i} 
$,
where $ [a_{i j}] \in \mathbb{R}^{N \times N}$ denotes the adjacency matrix of the subgraph $\G$ obtained from $\BG$ by removing node $0$ as well as those edges corresponding to node $0$, and $a_{i0}>0$ means that the $i$-th robot can receive the information from the leader. Then, it is reasonable to let $(v_i,w_i)$ rely only on $e_{\theta_{i}},e_{\tilde x_{i}}$ so as to reach the network constraints.

\section{Main Results}  \label{sec-Mresults}
	
The main results of this paper consist of three parts.
Firstly, a distributed continuous feedback control law is developed to solve Problem \ref{pro-01}. Secondly, the global $\mathcal{K}$-exponential stability of the closed-loop system is further proved under the proposed distributed continuous feedback control law. Finally, a distributed sampled-data feedback control law is synthesized based on the distributed continuous feedback control law by the emulation approach.

\subsection{Continuous Feedback Control Synthesis}\label{subsec-a}
	
Let us consider the following distributed continuous feedback law:
\begin{align}
	\omega_{i}=-k_{\omega}e_{\theta_{i}}+\omega_{0},~
	v_{i}=-k_{v}e_{\tilde x_{i}}+v_{0},~i=1,\dots,N  \label{cont-00}
\end{align} 
where $k_{\omega}$ and $k_{v}$ are positive constants. From the discussions in the previous section, it suffices to show \eqref{cont-00} stabilize system \eqref{error-sys}. Let $\bar x=[\bar x_{1}, \dots, \bar x_{N}]^{\t}$, $\bar y=[\bar y_{1}, \dots, \bar y_{N}]^{\t}$, and $\bar\theta=[\bar\theta_{1}, \dots, \bar\theta_{N}]^{\t}$. 
Then, the closed-loop system composed of \eqref{error-sys} and \eqref{cont-00} can be put into the compact form:
\begin{subequations}\label{CL-00}
	\begin{align}
		\dot {\bar\theta}&=-k_{\omega}H\bar \theta,\label{CL-01}\\
		\begin{bmatrix}	\dot{\bar x} \\ \dot{\bar y}	\end{bmatrix}&=
		\begin{bmatrix}	-k_{v}H & \omega_{0}I \\ -\omega_{0}I & 0	\end{bmatrix}
		\begin{bmatrix}	\bar x \\ \bar y	\end{bmatrix}+\begin{bmatrix}-k_{\omega}\tilde y_{0}H\bar\theta-k_{\omega}\mbox{diag}(\bar y)H\bar\theta  \\ k_{\omega}\tilde x_{0}H\bar\theta+k_{\omega}\mbox{diag}(\bar x)H\bar\theta\end{bmatrix}   \label{CL-02}
	\end{align}
\end{subequations}
where $H=[h_{ij}]\in \mathbb{R}^{N \times N}$ satisfies that $h_{ij}=-a_{ij}$ for any $i\neq j$, $i,j=1,\dots,N$, and $h_{ii}=\sum_{j=0}^Na_{ij}$, $i=1,\dots,N$.  
\begin{Theorem}\label{theorem-01}
Under Assumptions \ref{ass-02}-\ref{ass-04}, the origin of system \eqref{CL-00} is uniformly globally asymptotically stable (UGAS). Thus, Problem \ref{pro-01} is solved by \eqref{cont-00}.
\end{Theorem}
	
\begin{Proof}
The proof is based on finding the strictly Lyapunov function for system \eqref{CL-00}, which can be divided into  three steps:
		
\textit{Step-1: Show that the origin of \eqref{CL-01} is uniformly globally exponentially stable (UGES) by finding an explicit strict Lyapunov function.} By \cite[Lemma 1]{TAC-2012-SYF}, all the eigenvalues of the matrix $H$ have positive real parts under Assumption \ref{ass-04}. Thus, $H$ is a $\mathcal{M}$-matrix. By \cite[Theorem 4.25]{book-2009-QZH}, there exists a positive diagonal matrix $D=\mbox{diag}(d_{1},\dots, d_{N})$ such that $Q \triangleq DH+H^{\t}D$ is positive definite. Consider a positive definite and radially unbounded function
\begin{align}\label{V0}
	V_{0}=\frac{1}{2}\bar \theta^{\t}D\bar \theta.
\end{align}
Then, along the trajectory of subsystem \eqref{CL-01},
\begin{align}\label{der-V0}
	\dot V_{0}=-\frac{k_{\omega}}{2}\bar \theta^{\t}(DH+H^{\t}D)\bar\theta=-\frac{k_{\omega}}{2}\bar\theta^{\t} Q \bar \theta.
\end{align}
		
\textit{Step-2: Show that the origin of \eqref{CL-02} is iISS with respect to the input $\bar\theta(t)$ by finding an appropriate iISS-Lyapunov function.} Firstly, we define a Lyapunov function candidate
\begin{align}\label{SLF-V1}
	V_{1}=\frac{1}{2}(\bar x^{\t}D\bar x+\bar y^{\t}D\bar y).
\end{align} 
Along the trajectory of the nominal dynamics of \eqref{CL-02} (with $\bar\theta=0$), 
\begin{align*}
	&\dot V_{1} =\frac{1}{2}\bar x^{\t}D\left(-k_{v}H\bar x+\omega_{0}\bar y\right)+\frac{1}{2}\left(-k_{v}H\bar x+\omega_{0}\bar y\right)^{\t}D\bar x\notag\\
	&-\omega_{0}\bar y^{\t}D\bar x=-\frac{k_{v}}{2}\bar x^{\t}\left(DH+H^{\t}D\right)\bar x  \leq -\frac{k_{v}}{2}\lambda_{\min}(Q)\bar x^{\t}\bar x\leq 0.
\end{align*}
That is to say, $V_1$ is a weak Lyapunov function for the nominal dynamics of \eqref{CL-02}. Now, we can construct a iISS-Lyapunov function for system \eqref{CL-02}. The result is summarized by the following claim and its proof can be found in Appendix.
\begin{Claim}\label{lemma-slf}
\textit{Under Assumptions \ref{ass-02}-\ref{ass-04}. Let
\begin{align}\label{SLF-nominal}
	W_{1}(t,\bar x,\bar y)=(\varphi(t)+\gamma)V_{1}-\omega_{0}\bar x^{\t}D\bar y
\end{align}
with $V_1$ given in \eqref{SLF-V1} and 
\begin{align*}
	&\varphi(t)=1+\frac{2}{T}\int_{t-T}^{t}\int_{s}^{t}\omega_{0}^{2}(\tau){\rm d}\tau{\rm d}s,~\gamma
	\geq\max\Bigg\{\frac{2}{k_{v}\lambda_{\min}(Q)}\\
	&~~~~~~~~\Bigg[\frac{k_{v}\epsilon\bar\omega\Vert H \Vert^2}{2}+\frac{\epsilon\bar\omega\lambda_{\max}(D)}{2} +2\bar\omega^{2}\lambda_{\max}(D)\Bigg]-1,\bar \omega\Bigg\}. 
\end{align*}
Then, along the trajectory of subsystem \eqref{CL-02},
\begin{align}\label{der-w1-02}
	\dot W_{1} 
	&\leq -\frac{\mu}{T}V_{1}+\left(C_{1}V_{1}+C_{2}\sqrt{V_{1}}\right)\Vert \bar\theta \Vert
\end{align}
for some $C_1,C_2>0$.}
\end{Claim}
		
Now, we define 
\begin{align}\label{Weq}
	W_3=\int_{0}^{W_{2}}\left(\frac{e^{s}-1}{e^{s}}\right)\mathrm{d}s  \mbox{~with~} W_{2}=\ln(1+W_{1}).
\end{align}
From \eqref{der-w1-02}, along the trajectory of subsystem \eqref{CL-02},
\begin{align}
	&	\dot W_{2}
	\leq -\frac{\mu V_{1}}{T(1+W_{1})}+\frac{C_{1}V_{1}+C_{2}\sqrt{V_{1}}}{1+W_{1}}\left\Vert\bar\theta\right\Vert\notag\\
	&\leq -\frac{\mu}{T\left(1+T\bar\omega+2\gamma\right)}\frac{W_{1}}{
	1+W_{1}}+\frac{C_{1}W_{1}+C_{2}\sqrt{W_{1}}}{1+W_{1}}\left\Vert\bar\theta\right\Vert. \label{der-w2-02bbb}
\end{align}
With $C_{0}=\max_{\sqrt{W_{1}}\geq 0}\frac{C_{1}W_{1}+C_{2}\sqrt{W_{1}}}{1+W_{1}}>0$ and $\overline C_{0}=\frac{\mu}{T\left(1+T\bar\omega+2\gamma\right)}>0$, from \eqref{der-w2-02bbb}, along the trajectory of \eqref{CL-02},  
\begin{align}
	\dot W_3&=\frac{e^{W_{2}}-1}{e^{W_{2}}}\dot W_{2} \leq \left(\frac{W_{1}}{1+W_{1}}\right)\left(-\overline C_{0}\frac{W_{1}}{1+W_{1}}+C_{0}\Vert \bar \theta \Vert \right)\notag\\
	& \leq -\frac{\overline C_{0}}{2}\left(\frac{V_{1}}{1+V_{1}}\right)^2+\frac{C^2_{0}}{2\overline C_{0}}\Vert \bar\theta \Vert^2.  \label{eq-W-uuu}
\end{align}

\textit{Step 3: Show the origin of \eqref{CL-00} is UGAS by jointing Lyapunov functions \eqref{V0} and \eqref{Weq} together.} Let us consider  
\begin{align}\label{omega}
	\Omega=W_3+\sigma V_{0}.
\end{align}
Then, from \eqref{der-V0} and \eqref{eq-W-uuu}, with $\frac{\lambda_{\min}(Q)}{2}\sigma k_{\omega}\geq \frac{C^2_{0}}{\overline C_{0}},$ i.e., $\sigma \geq \frac{2C^2_{0}}{\lambda_{\min}(Q)k_{\omega}\overline C_{0}}$, along the  trajectory of \eqref{CL-00},
\begin{align}\label{der-omega}
	\dot\Omega
	&\leq  -\frac{\overline C_{0}}{2} \left(\frac{V_{1}}{1+V_{1}}\right)^2-\frac{C^2_{0}}{2\overline C_{0}}\Vert \bar\theta \Vert^2.
\end{align}
Thus, $\Omega$ is an appropriate  strict Lyapunov function for system \eqref{CL-00}. 
\end{Proof}

\begin{Remark}
The idea of constructing $W_1$ by \eqref{SLF-nominal} is inspired by \cite{book-2009-Mazenc} and \cite{Auto-2017-Maghenem}. \cite[Chapter 6]{book-2009-Mazenc} provides a systematic framework on constructing time-varying strict Lyapunov function from weak Lyapunov function together with PE signals. Reference \cite{Auto-2017-Maghenem} refines this process for a class of linear time-varying systems. The nominal dynamics of \eqref{CL-02} here have a similar structure to the ``adaptive control system" in \cite{Auto-2017-Maghenem}. However, the strict Lyapunov function in \cite{Auto-2017-Maghenem} can only work when $D$ is the identity matrix. While, we consider here a more general case where $D$ is a positive definite diagonal matrix.
\end{Remark}

\begin{Remark}
The idea of constructing $W_3$ by \eqref{Weq} is motivated by the ``changing supply function" technique for iISS-Lyapunov function proposed by \cite{TAC-2010-Ito}. The main purpose is to change $ \Vert\bar\theta \Vert$ in \eqref{der-w1-02} to $ \Vert\bar\theta \Vert^2$ in \eqref{eq-W-uuu} so as to match the form of \eqref{der-V0}. 
\end{Remark}

\begin{Remark}
Inequality \eqref{eq-W-uuu} means \eqref{CL-02} is iISS, see \cite[Definition 2.1]{book-2009-Mazenc}, and \eqref{der-w1-02} further implies that \eqref{CL-02} is small input-to-state stable in the sense that 
$
	\Vert \bar \theta \Vert \leq \min\{\frac{\mu}{4TC_{1}}, \frac{\mu}{4TC_{2}}\sqrt{V_{1}}\}$ implies that $\dot W_{1}\leq -\frac{\mu}{2T}V_{1}.
$
From \cite[Eq. (12)]{TAC-2014-Ito}, these two properties together indicate that \eqref{CL-02} is indeed strongly iISS  with respect to $\Vert \bar \theta \Vert$, see \cite[Definition 2]{TAC-2014-Ito}. Notice that the key idea that the input is limited by some certain value will be used later to analyze the convergence of the sampled-data control system.
\end{Remark}
	

\subsection{Exponential Convergence Rate of the Closed-Loop System}\label{subsec-b}
	
We further investigate the convergence rate of system \eqref{CL-00}  and show the global $\mathcal{K}$-exponential stability. 
\begin{Theorem}\label{theorem-02}
Under Assumptions \ref{ass-02}-\ref{ass-04}, with $C_1$ and $C_2$, $C_0$ and $\overline C_0$, and $\sigma$ given in \eqref{der-w1-02}, \eqref{eq-W-uuu}, and \eqref{omega} respectively, the origin of \eqref{CL-00} is globally $\mathcal{K}$-exponentially stable, in the sense that, for any initial state $\Vert[\bar x^{\t}(t_{0}), \bar y^{\t}(t_{0}), \bar\theta^{\t}(t_{0})]\Vert\triangleq r_{0}$, the solution of \eqref{CL-00} satisfies
\begin{align}
	&\Vert [\bar x^{\t}(t), \bar y^{\t}(t), \bar\theta^{\t}(t)] \Vert
	\leq  M_0(r_0) e^{- {C_{3}} (t-t_{0})}  \label{eq-Kexp}
\end{align}
for some class $\mathcal{K}$ function $M_0(\cdot)$ and some positive constant $C_3$ (independent of $r_0$).
\end{Theorem}

\begin{Proof}
At first, for any initial value $\Vert[\bar x^{\t}(t_{0}), \bar y^{\t}(t_{0}), \bar\theta^{\t}(t_{0})]\Vert\triangleq r_{0}$, we give an explicit class $\mathcal{K}$ function $\delta(\cdot)$, such that $\Vert [\bar x^{\t}(t), \bar y^{\t}(t), \bar\theta^{\t}(t)] \Vert \leq \delta(r_{0})$.  
Combining \eqref{omega} and \eqref{der-omega}, we have
$
\Omega(t)\leq \Omega(t_{0}),~\forall t\geq t_{0}.
$ With some calculations, 
\begin{align}\label{ieq-oemga}
	&\Omega(t)
	=\ln(1+W_{1})-\frac{W_{1}}{1+W_{1}}+\sigma V_{0}\notag\\
	&\leq \ln(1+W_{1})+\frac{\sigma}{2}\bar \theta^{\t}D\bar \theta
	\leq \frac{\mu}{T\overline C_0}V_{1}+\frac{\sigma}{2}\bar \theta^{\t}D\bar \theta\notag\\
	&\leq \frac{\lambda_{\max}(D)}{2}\max\{\mu/(T\overline C_0), \sigma\}\Vert [\bar x^{\t}(t), \bar y^{\t}(t), \bar\theta^{\t}(t)] \Vert^2
\end{align} 
which implies
$
\Omega(t_{0})\leq \frac{\lambda_{\max}(D)}{2}\max\{\mu/T\overline C_0, \sigma\}r^2_{0}.
$
		
To construct an explicit $\delta(\cdot)$, we consider the following two cases:

Case-1: $0\leq W_{1} \leq 7$. In this case,  
$
\frac{1}{128}W^2_{1}\leq \ln(1+W_{1})-\frac{W_{1}}{1+W_{1}}
$
which implies 
$	\frac{1}{128}V^2_{1}\leq \frac{1}{128}W^2_{1}+\frac{\sigma\lambda_{\min}(D)}{2}\Vert \bar \theta \Vert^2\leq \Omega.$
This together with \eqref{ieq-oemga} gives
\begin{align}\label{add-3.2-01}
	\Vert [\bar x^{\t}(t), \bar y^{\t}(t)]\Vert^2\leq \frac{16\sqrt{\lambda_{\max}(D)\max\{\mu/(T\overline C_0), \sigma\}}}{\lambda_{\min}(D)}r_{0}.
\end{align}
By \eqref{V0} and \eqref{der-V0}, for any $t\geq t_{0}$, 
$
V_{0}(t)\leq V_{0}(t_{0})
$,
which implies
\begin{align}\label{add-3.2-02}
	\Vert \bar \theta(t) \Vert^2\leq \frac{\lambda_{\max}(D)}{\lambda_{\min}(D)}\Vert \bar \theta(t_{0}) \Vert^2\leq \frac{\lambda_{\max}(D)}{\lambda_{\min}(D)}r^2_{0}.
\end{align}
Thus, by \eqref{add-3.2-01} and \eqref{add-3.2-02},  
$	\Vert [\bar x^{\t}(t), \bar y^{\t}(t), \bar\theta^{\t}(t)] \Vert
\leq \delta_{1}(r_{0})$
where 
$
\delta_{1}(r_{0})=\sqrt{\frac{\lambda_{\max}(D)}{\lambda_{\min}(D)}r^2_{0}+\frac{16\sqrt{\lambda_{\max}(D)\max\{\mu/(T\overline C_0), \sigma\}}}{\lambda_{\min}(D)}r_{0}}
$.
		
Case-2: $7\leq W_{1}$. In this case, 
$
\frac{1}{2}\ln(1+W_{1})\leq \ln(1+W_{1})-\frac{W_{1}}{1+W_{1}}
$.
Since 
$
\ln(1+\sigma V_{0})\leq \sigma V_{0}
$,
we obtain 
\begin{align}\label{ieq-wi-case2}
	&\ln(1+W_{1})-\frac{W_{1}}{1+W_{1}}+\sigma V_{0}\notag\\
	&\geq \frac{1}{2}(\ln(1+W_{1})+\ln(1+\sigma V_{0}))\geq \frac{1}{2}\ln(1+W_{1}+\sigma V_{0})\notag\\
	&\geq \frac{1}{2}\ln(1+\frac{\lambda_{\min}(D)}{2}\min\{1, \sigma\}\Vert [\bar x^{\t}(t), \bar y^{\t}(t), \bar\theta^{\t}(t)] \Vert^2 ). 
\end{align}
By \eqref{ieq-oemga} and \eqref{ieq-wi-case2},  
\begin{align*}
	&\frac{1}{2}\ln(1+\frac{\lambda_{\min}(D)}{2}\min\{1, \sigma\}\Vert [\bar x^{\t}(t), \bar y^{\t}(t), \bar\theta^{\t}(t)] \Vert^2 )\notag\\
	&\leq \frac{1}{2}\ln(1+W_{1}+\sigma V_{0})\leq \Omega(t)\leq \Omega(t_{0})\notag\\
	&\leq \frac{\lambda_{\max}(D)}{2}\max\{(\mu/(T\overline C_0), \sigma\}r^2_{0}.
\end{align*}
Thus,  
$
\Vert [\bar x^{\t}(t), \bar y^{\t}(t), \bar\theta^{\t}(t)] \Vert
\leq \delta_{2}(r_{0})
$
where 
$
\delta_{2}(r_{0})=\sqrt{\frac{2e^{\lambda_{\max}(D)\max\{\mu/(T\overline C_0), \sigma\}r^2_{0}}-2}{\lambda_{\min}(D)\min\{1, \sigma\}}}
$. Now, letting $\delta(s)=\max\{\delta_{1}(s), \delta_{2}(s)\}$ for $s\geq 0$ gives
$
\Vert (\bar x^{\t}(t), \bar y^{\t}(t), \bar\theta^{\t}(t)) \Vert \leq \delta(r_{0})
$.
		
Next, we show the state of the closed-loop system converges exponentially. Notice that $\sqrt{V_{1}}\leq \sqrt{\lambda_{\max}(D)/2}\delta(r_{0})$. From \eqref{der-w1-02}, 
\begin{align*}
	\dot W_{1}
	&\leq -\frac{\mu}{T}V_{1}+\left(C_{1}V_{1}+C_{2}\sqrt{V_{1}}\right)\left\Vert \bar\theta \right\Vert\notag\\
	&\leq -\frac{\mu}{T}V_{1}+\bar \delta(r_{0})\sqrt{V_{1}}\left\Vert \bar\theta \right\Vert 
	\leq -\frac{\mu}{2T}V_{1}+\frac{T}{2\mu}\bar\delta^{2}(r_{0})\left\Vert \bar\theta \right\Vert^{2}
\end{align*}
where $\bar\delta(r_{0})=\left(C_{1}\sqrt{\lambda_{\max}(D)/2}\delta(r_{0})+C_{2}\right)$. Let $W_{4}=W_{1}+\Delta_{0} V_{0}$ with $\Delta_{0}=\frac{2T}{\mu k_{\omega}\lambda_{\min}(Q)}\bar\delta^{2}(r_0)$. Then, along the trajectory of the closed-loop system \eqref{CL-00},
\begin{align*}
	\dot W_{4}&\leq -\frac{\mu}{2T}V_{1}+\frac{T}{2\mu}\bar\delta^{2}(r_{0})\left\Vert \bar\theta \right\Vert^{2}-\frac{k_{\omega}}{2}\lambda_{\min}(Q)\Delta_{0}\Vert \bar \theta\Vert ^{2}\notag\\
	&\leq -\frac{\mu}{2T}V_{1}-\frac{k_{\omega}}{4}\lambda_{\min}(Q)\Delta_{0}\Vert \bar \theta\Vert ^{2}\notag\\
	&\leq -\frac{\overline C_{0}}{2}W_{1}-\frac{k_{\omega}\lambda_{\min}(Q)}{2\lambda_{\max}(D)}\Delta_{0}V_{0} \leq -2C_{3}W_{4}
\end{align*}
where $C_{3}=\min\left\{\frac{\overline C_{0}}{4}, \frac{k_{\omega}\lambda_{\min}(Q)}{4\lambda_{\max}(D)}\right\}$. Then, it follows that
\begin{align*}
	W_{4}(t,\bar  x(t), \bar y(t), \bar\theta(t))\leq W_{4}(t_{0},\bar  x(t_{0}), \bar y(t_{0}), \bar\theta(t_{0}))e^{-2C_{3}(t-t_{0})}.
\end{align*}
According to \eqref{V0}, \eqref{SLF-nominal}, and \eqref{ieq-w1},  
\begin{align*}
	&W_{4}(t_{0},\bar  x(t_{0}), \bar y(t_{0}), \bar\theta(t_{0}))
	\leq  \frac{\mu}{T\overline C_{0}}V_{1}(\bar x(t_{0}), \bar y(t_{0}))+\Delta_{0}V_{0}(\theta(t_0))\notag\\
	&~~~~~~~~~~~~~~~~~~~~~~~~~~~~~~~\leq  \frac{\lambda_{\max}(D)\max\{\mu/T\overline C_{0}, \Delta_{0}\}}{2}r^2_{0}, \notag\\
	&W_{4}(t,\bar  x(t), \bar y(t), \bar\theta(t))
	\geq V_{1}(\bar x(t), \bar y(t))+\Delta_{0}V_{0}(\theta(t))\geq  \frac{\lambda_{\min}(D)}{2}  \notag\\
	&~~~~~\times \min\{1, 2TC^2_{2}/(\mu k_{\omega}\lambda_{\min}(Q))\}\Vert[\bar x^{\t}(t), \bar y^{\t}(t), \bar\theta^{\t}(t)] \Vert^2 .
\end{align*}
Letting $M_0(r_0)=\sqrt{\frac{\max\{\mu/T\overline C_0, \Delta_{0}\}\lambda_{\max}(D)}{\min\{1, 2TC^2_{2}/(\mu k_{\omega}\lambda_{\min}(Q))\}\lambda_{\min}(D)}}r_{0}$ yields \eqref{eq-Kexp}.  
\end{Proof}

\begin{Remark}
The choice of the number $7$ in the proof is just for convenience. The idea behind this choice is to find appropriate positive numbers $c_{0}$, $c_{1}$, and $b $ such that the following hold
\begin{align*}
		\left\{
	\begin{aligned}
		c_{0}\ln(1+W_{1}) \leq & \ln(1+W_{1})-\frac{W_{1}}{1+W_{1}},~   &W_{1}\geq b , \\ 
		c_{1}W^2_{1} \leq &\ln(1+W_{1})-\frac{W_{1}}{1+W_{1}},~  &0 \leq W_{1}\leq b . 
	\end{aligned} \right.
\end{align*}
Clearly, there are many combinations of valid numbers $(c_{0},c_{1},b)$, for example $(\frac{1}{2},\frac{1}{128},7)$ here.
\end{Remark}

\begin{Remark}
In \cite{TAC-2018-Loria}, a formation-tracking control law was developed for mobile robots, and uniform global asymptotic stability of the closed-loop system was achieved by constructing strict/iISS Lyapunov functions for subsystems provided that the communication network is a pure spanning tree. The result of this paper improves that of \cite{TAC-2018-Loria} from three aspects. First, we allow more general directed communication networks which suffice to contain spanning trees. Second, an explicit strict Lyapunov function is constructed for the whole closed-loop system. Third, the exponential convergence rate is further characterized and it is proven that the closed-loop system is globally $\mathcal{K}$-exponentially stable. 
\end{Remark}

\subsection{Sampled-Data Feedback Control  Synthesis}\label{sub-033}
We further consider the sampled-data control for Problem \ref{pro-01}. Motivated by \eqref{cont-00}, we design the following distributed sampled-data feedback control law: for $i=1,\dots,N$,
\begin{subequations}\label{cont-000}
	\begin{align}
		\omega_{i}(t)&=-k_{\omega}e_{\theta_{i}}(t_{k})+\omega_{0}(t), \label{samp-cont-001}\\
		v_{i}(t)&=-k_{v}e_{\tilde x_{i}}(t_{k})+v_{0}(t),~\forall t\in[t_{k}, t_{k+1})\label{samp-cont-002}
	\end{align}
\end{subequations} 
where $t_{k}$ is the sampling instant, $t_{k+1}-t_{k}=T_{0}$, and $e_{\theta_{i}}(t_{k})=\sum_{j=0}^{N}a_{ij}(\theta_{i}(t_k)-\theta_{j}(t_{k})),$ $e_{\tilde x_{i}}(t_{k})=\sum_{j=0}^{N}a_{ij}(\tilde x_{i}(t_k)-\tilde x_{j}(t_k))$.
Then, the closed-loop system composed of \eqref{error-sys} and \eqref{cont-000} can be put into the compact form
\begin{subequations}\label{SP-CL-00}
	\begin{align}
		\dot {\bar\theta}(t)&=-k_{\omega}H\bar \theta(t_{k})=-k_{\omega}H\bar \theta(t)-k_{\omega}H\hat\theta(t),\label{SP-CL-01}\\
		\begin{bmatrix}	\dot{\bar x}(t) \\ \dot{\bar y}(t)	\end{bmatrix}&=
		\begin{bmatrix}	-k_{v}H & \omega_{0}I \\ -\omega_{0}I & 0	\end{bmatrix}
		\begin{bmatrix}	\bar x(t) \\ \bar y(t)	\end{bmatrix}+
		\begin{bmatrix}	-k_{\omega}\mbox{diag}(\bar y)H\bar\theta(t_{k}) \\ k_{\omega}\mbox{diag}(\bar x)H\bar\theta(t_{k})\end{bmatrix}\notag\\
		&~~~~+\begin{bmatrix}-k_{\omega}\tilde y_{0}H\bar\theta(t_{k}) \\ k_{\omega}\tilde x_{0}H\bar\theta(t_{k})\end{bmatrix}+\begin{bmatrix}	-k_{v}H\hat x(t) \\ 0 \end{bmatrix} \label{SP-CL-02}
	\end{align}
\end{subequations}
for $t\in[t_{k}, t_{k+1})$, where $\hat\theta(t)=\bar\theta(t_{k})-\bar\theta(t)$ and $\hat x(t)=\bar x(t_{k})-x(t)$.

\begin{Theorem}\label{theorem-03}
Under Assumptions \ref{ass-02}-\ref{ass-04}, there exists $T^*>0$, such that the origin of \eqref{SP-CL-00} is UGAS if the sampling period $T_{0}$ satisfies $0<T_{0}< T^{*}$. Thus, Problem \ref{pro-01} is solved by \eqref{cont-000}.  
\end{Theorem}
	
\begin{Proof}
For simplicity, we denote $V_0(t) \triangleq V_0(\bar{\theta}(t)) $, $V_1(t) \triangleq V_1(\bar x(t),\bar y(t))$, $W_{1}(t) \triangleq W_{1}(t, \bar x(t), \bar y(t)) $, where $V_0$, $V_1$, and $W_1$ are given in \eqref{V0}, \eqref{SLF-V1}, and \eqref{der-w1-02}, respectively. The proof can be divided into three steps:
		
\textit{Step 1: Show that there exists $T_1^*>0$ such that the origin of \eqref{SP-CL-01} is UGAS if $0<T_{0}< T^{*}_1$.}
Note that, if $\bar \theta(t_{k})=0$ at some sampling instant $t_{k}$, then, from \eqref{SP-CL-01}, $\bar \theta(t)=0$ for all $t\geq t_{k}$. In this case, the origin of \eqref{SP-CL-01} is UGAS. Now, suppose that $\bar \theta(t_{k})\neq 0$ at any sampling instant $t_k$. Then, along the trajectory of \eqref{SP-CL-01},  
\begin{align}\label{sp-v0}
	\dot V_{0}(t)
	&\leq -\frac{k_{\omega}\lambda_{\min}(Q)}{\lambda_{\max}(D)}V_{0}(t)+k_{\omega}\Vert \bar\theta(t) \Vert \Vert DH \Vert \Vert \hat\theta(t)\Vert.
\end{align}
On the other hand,  according to \eqref{SP-CL-01}, for any $t\in[t_{k}, t_{k+1})$,
\begin{align*}
	\Vert\dot{\bar\theta}(t)\Vert
	&=\Vert k_{\omega}H\bar\theta(t_{k})\Vert
	\leq \frac{k_{\omega}\Vert H \Vert}{\sqrt{\lambda_{\min}(D)}}\sqrt{2V_{0M}(t)}  
\end{align*}
where $V_{0M}(t)=\max_{\tau\in[t_{k}, t]}V_{0}(\tau)$. Then, it follows
\begin{align}\label{sp-theta}
	\Vert \hat\theta(t) \Vert
	\leq \int_{t_{k}}^{t}\Vert
	\dot{\bar\theta}(\tau)\Vert{\rm d}\tau \leq \frac{k_{\omega}T_{0}\Vert H \Vert}{\sqrt{\lambda_{\min}(D)}}\sqrt{2V_{0M}(t)} . 
\end{align}
Combining \eqref{sp-v0} and \eqref{sp-theta}, for any $t\in[t_{k}, t_{k+1})$, we have
\begin{align}\label{sp-v0-2}
	\dot V_{0}(t)
    &\leq -h_{1}V_{0}(t)+h_{2}T_{0}\sqrt{V_{0}(t)}\sqrt{V_{0M}(t)}
\end{align}
where $h_{1}=\frac{k_{\omega}\lambda_{\min}(Q)}{\lambda_{\max}(D)}$, $h_{2}= \frac{2k^2_{\omega}\Vert H \Vert \Vert DH \Vert}{\lambda_{\min}(D)}$. Let $T_1^*=\frac{h_{1}}{h_{2}}$. We claim that, whenever $0<T_{0}< T^{*}_1$,
\begin{align}\label{sp-max-v0}
	\sqrt{V_{0M}(t)}=\sqrt{V_{0}(t_{k})},\ \forall t\in[t_{k}, t_{k+1}).
\end{align}
Otherwise, there exists time instant  $t^{*}\in[t_{k}, t_{k+1})$ such that $V_{0}(t^{*})>V_{0}(t_{k})$. From \eqref{sp-v0}, 
$\dot V_{0}(t_{k})<0$ for any $\bar\theta(t_{k})\neq 0$. 
There exists another time instant $t^{**}\in[t_{k}, t^{*})$, such that
\begin{align*}
	V_{0}(t^{**})=V(t_{k}),~  
	\dot V_{0}(t^{**})>0,~
	V_{0}(t)\leq V_{0}(t^{**}),~\forall t\in[t_{k}, t^{*}] . 
\end{align*} 
This together with \eqref{sp-v0-2} gives $\dot V_{0}(t^{**}) \leq -h_{1}V_{0}(t^{**})+h_{2}T_{0}V_{0}(t^{**})$. Then, it holds also $\dot V_{0}(t^{**})<0$ by the choice of $T_0$, reaching a contradiction. So \eqref{sp-max-v0} is true.

Let $\xi(t)=\sqrt{\frac{V_{0}(t)}{V_{0}(t_{k})}}$ for any $t\in[t_{k}, t_{k+1})$. Then, \eqref{sp-v0-2} and \eqref{sp-max-v0} imply
$
\dot \xi(t)\leq -\frac{h_{1}}{2}\xi(t)+\frac{h_{2}}{2}T_{0}
$.
Since $\xi(t_{k})=1$, we obtain
\begin{align}\label{sp-rho}
	\xi(t_{k+1})& \leq 
	e^{-\frac{h_{1}}{2}T_{0}}(1-\frac{h_{2}T_{0}}{h_{1}})+\frac{h_{2}T_{0}}{h_{1}}:=\varrho
\end{align}
which implies  $V_{0}(t_{k+1})\leq \varrho^2 V_{0}(t_{k}).$ Again by the choice of $T_0$, from \eqref{sp-rho}, we have $0<\varrho<1$. Then, it follows that $V_{0}(t_{k})$ converges to zero. By \eqref{sp-max-v0}, $V_{0M}(t)$, hence $V_{0}(t)$, converges to zero, and so does $ \bar\theta(t) $. Again, from \eqref{sp-max-v0}, we know the origin of \eqref{SP-CL-01} is uniformly globally stable (UGS), hence, it is UGAS.

\textit{Step 2: Show that there exists $T_2^*>0$ such that the origin of \eqref{SP-CL-02} is uniformly globally attractive (UGA) and the state of \eqref{SP-CL-02} is uniformly globally bounded (UGB) if $T_{0}$ satisfies $0<T_{0}< \min\{T^{*}_1,T^{*}_2\} \triangleq T^{*}$.} Following the same argument as the proof of Claim \ref{lemma-slf}, we have $W_{1}$ satisfies,  along the trajectory of \eqref{SP-CL-02},
\begin{align}\label{sp-der-w1}
	\dot W_{1}(t)
	&\leq -\frac{\mu}{T}V_{1}(t)+\left(C_{1}V_{1}(t)+C_{2}\sqrt{V_{1}(t)}\right)\Vert \bar\theta(t_{k}) \Vert\notag\\
	&~~~~+C_{4}\sqrt{V_{1}(t)}\Vert \hat x(t) \Vert\notag\\
	&\leq -\overline{C}_{0}W_{1}(t)+\left(C_{1}W_{1}(t)+C_{2}\sqrt{W_{1}(t)}\right)\Vert \bar\theta(t_{k}) \Vert\notag\\
	&~~~~+C_{4}\sqrt{W_{1}(t)}\Vert \hat x(t) \Vert,\forall t\in[t_{k}, t_{k+1})
\end{align}
where $\overline{C}_{0}$ is defined as \eqref{eq-W-uuu}, $C_{1}$, $C_{2}$ are the same as those in \eqref{der-w1-02}, and  $C_{4}=2k_{v}(1+\gamma+T\bar\omega+\bar\omega)\sqrt{\lambda_{\max}(D)}\Vert H \Vert$. 
We first claim that the state $[\bar x^{\t}, \bar y^{\t}]^{\t}$ of \eqref{SP-CL-02} does not escape in any finite time.
Combining \eqref{SLF-V1} and \eqref{SP-CL-02}, for any $t\in[t_{k}, t_{k+1})$, 
\begin{align*}
	\dot V_{1}(t)
	&\leq k_{v}\Vert \bar x^{\t}(t)D\Vert \Vert H\bar x(t_{k})\Vert \notag\\
	&~~~+\sqrt{2}k_{\omega}M(\Vert \bar x^{\t}(t) D \Vert + \Vert \bar y^{\t} D \Vert)\Vert H\bar\theta(t_{k})\Vert\notag\\
	&\leq 2k_{v}\Vert H \Vert\sqrt{\frac{\lambda_{\max}(D)}{\lambda_{\min}(D)}}\sqrt{V_{1}(t)}\sqrt{V_{1}(t_{k})}\notag\\
	&~~~+4k_{\omega}M\Vert H \Vert\sqrt{\frac{\lambda_{\max}(D)}{\lambda_{\min}(D)}}\sqrt{V_{1}(t)}\sqrt{V_{0}(t_{k})}.
\end{align*}
Let $L_{1}=k_{v}\Vert H \Vert\sqrt{\frac{\lambda_{\max}(D)}{\lambda_{\min}(D)}}$ and $L_{2}=2k_{\omega}M\Vert H \Vert\sqrt{\frac{\lambda_{\max}(D)}{\lambda_{\min}(D)}}$. Then, it follows that
$
\dot {\sqrt{V_{1}(t)}}\leq L_{1}\sqrt{V_{1}(t_{k})}+L_{2}\sqrt{V_{0}(t_{k})}, 
$
so
\begin{align}\label{sp-sv1}
	\sqrt{V_{1}(t)}
	&\leq (1+T_{0}L_{1})\sqrt{V_{1}(t_{k})} +T_{0}L_{2}\sqrt{V_{0}(t_{k})}
\end{align}
for any $t\in[t_{k}, t_{k+1})$. Consequently,  for any  $t_{0}\leq t <\infty$, 
\begin{align}\label{finite time-v1}
	\sqrt{V_{1}(t )}
	&\leq (1+T_{0}L_{1})^{\lfloor\frac{t -t_{0}}{T_{0}}\rfloor+1}\sqrt{V_{1}(t_{0})}+\frac{L_{2}}{L_{1}} \notag\\
	&~~~~\times\left((1+T_{0}L_{1})^{\lfloor\frac{t -t_{0}}{T_{0}}\rfloor+1}-1\right) \sqrt{V_{0}(t_{0})}
\end{align}
where $V_{0}(t_{k+1})\leq \varrho^2 V_{0}(t_{k}) < V_{0}(t_{k}) $ is used. The claim is true.
		
Next, let 
$$
T^*_2=\min\left\{\frac{3\overline{C}_{0}}{4C_{2}}, \frac{-C_{2}-C_{4}L_{3}+\sqrt{(C_{2}+C_{4}L_{3})^2+3\overline{C}_{0}C_4C_{5} }}{2C_4C_{5}} \right\}. 	 
$$
\noindent We show that the origin of \eqref{SP-CL-02} is UGA whenever $0<T_{0}< \min\{T^{*}_1,T^{*}_2\} \triangleq T^{*}$, consequently, by continuity, the state of \eqref{SP-CL-02} is UGB. Since the origin of 
\eqref{SP-CL-01}  is UGAS, there exists a sampling instant $t_{k_0}$, $t_{0}\leq t_{k_0}<\infty$, such that $\Vert\bar \theta({t}) \Vert\leq \frac{\overline{C}_{0}}{4C_{1}}$ for $t\in[t_{k_0},\infty)$. Then, \eqref{sp-der-w1} implies, for $t\in[t_{k_0},\infty)$, 
\begin{align}\label{add-dotW1}
	\dot W_{1}(t)\leq -\frac{3}{4}\overline{C}_{0}W_{1}(t)+(C_{2}\Vert \bar\theta(t_{k}) \Vert+C_{4}\Vert \hat x(t) \Vert)\sqrt{W_{1}(t)} . 
\end{align} 
For any $k\geq k_0$, two cases may happen:
		
Case-1: $\Vert \bar \theta(t_k)  \Vert\leq T_{0}\sqrt{W_{1}(t_k)}$. For $t\in[t_{k}, t_{k+1})$, from \eqref{SP-CL-02},  
\begin{align}\label{add-boundx}
	\Vert \dot{\bar x}(t) \Vert
		&\leq {k_{v}\Vert H\bar x(t_{k}) \Vert}+\bar w\Vert \bar y(t) \Vert \notag\\
		&~~~~+k_{\omega}\Vert \bar y(t) \Vert \Vert H\bar\theta(t_{k})\Vert + \sqrt{2}k_{\omega}M\Vert H\bar\theta(t_{k})\Vert\notag\\
		&\leq \left(L_{3}\sqrt{W_{1M}(t)}+C_{5}T_{0} \sqrt{W_{1M}(t)}\right)
\end{align}
where $L_{3}=\sqrt{\frac{2}{\lambda_{\min}(D)}}(k_{v}\Vert H \Vert+\frac{ k_{\omega}\overline C_{0}}{4C_{1}}\Vert H \Vert+\bar\omega)$, $C_{5}=\sqrt{2}k_{\omega}M\Vert H \Vert$, and $W_{1M}(t)=\max_{\tau\in[t_{k}, t]}W_{1}(\tau)$.
It follows that
\begin{align*}
	\Vert \hat x(t) \Vert&\leq \int_{t_{k}}^{t}\Vert \dot{\bar x}(\tau) \Vert{\rm d}\tau\leq \left(L_{3}\sqrt{W_{1M}(t)}+C_{5}T_{0}\sqrt{W_{1M}(t)}\right)T_{0}.
\end{align*}
Furthermore, 
\begin{align}\label{3.3-w1}
	\dot W_{1}(t)
	&\leq -\frac{3}{4}\overline{C}_{0}W_{1}(t)+C_{6}\sqrt{W_{1}(t)}\sqrt{W_{1M}(t)} 
\end{align}
where $C_{6}=C_{4}C_{5}T^{2}_{0}+(C_{4}L_{3}+C_{2})T_{0}$.
  On the other hand, notice that $\Vert \hat x(t) \Vert =0$ for all $t_{k}$. Hence, from \eqref{add-dotW1}, $\dot W_{1}$ at $t_{k}$ satisfies
\begin{align*}
	\dot W_{1}(t_{k})
		&\leq -\frac{3}{4}\overline{C}_{0}W_{1}(t_{k})+C_{2}\sqrt{W_{1}(t_{k})}\left\Vert \bar\theta(t_{k}) \right\Vert\\
		&\leq  -\frac{3}{4}\overline{C}_{0}W_{1}(t_{k})+T_{0}C_{2}W_{1}(t_{k}). 
\end{align*}
Whenever $T_{0}<  \frac{3\overline{C}_{0}}{4C_{2}}$, $\dot W_{1}(t_{k})< 0$ for any $\Vert [\bar x^{\t}(t_{k}), \bar y^{\t}(t_{k})] \Vert\neq 0$. Now, with the same argument as that of \eqref{sp-max-v0}, we can claim that, whenever $0<T_{0}<T^{*}_{2}$,
\begin{align}\label{sp-max2}
	\sqrt{W_{1M}(t)}=\sqrt{W_{1}(t_{k})},\ \forall t\in[t_{k}, t_{k+1}) .
\end{align}
Let $\zeta(t)=\sqrt{\frac{W_{1}(t)}{W_{1}(t_{k})}}$ for any $t\in[t_{k}, t_{k+1})$.  \eqref{3.3-w1} can be transformed to 
$
\dot \zeta(t)\leq -\frac{3}{8}\overline{C}_{0}\zeta(t)+\frac{C_{6}}{2}
$.
With the same argument as \eqref{sp-rho}, 
\begin{align}\label{eq-chi}
	0<\zeta(t_{k+1})=e^{-\frac{3}{8}\overline{C}_{0}T_{0}}(1-\frac{4C_{6}}{3\overline{C}_{0}})+\frac{4C_{6}}{3\overline{C}_{0}}:=\chi<1
\end{align}
which implies 
$
\sqrt{W_{1}(t_{k+1})}\leq \chi  \sqrt{W_{1}(t_{k})}
$.  
		
Case-2: $\Vert\bar \theta({t_k}) \Vert> T_{0}\sqrt{W_{1}(t_k)}$. By \eqref{sp-sv1} and \eqref{ieq-w1} in Appendix, for $t\in[t_{k}, t_{k+1})$,
\begin{align}
	\sqrt{W_{1}(t)}
	&\leq \sqrt{(1+T\bar\omega+2\gamma)}\sqrt{V_{1}(t)}\notag\\
	&\leq \sqrt{\frac{\mu}{T\overline{C}_{0}}}\left((1+T_{0}L_{1})\sqrt{V_{1}(t_{k})} +T_{0}L_{2}\sqrt{V_{0}(t_{k})}\right)\notag\\
	&\leq \sqrt{\frac{\mu}{T\overline{C}_{0}}}\left((1+T_{0}L_{1})\sqrt{W_{1}(t_{k})} +T_{0}L_{2}\sqrt{V_{0}(t_{k})}\right)\notag\\
	&\leq \sqrt{\frac{\mu}{T\overline{C}_{0}}}\left((\frac{1}{T_{0}}+L_{1})\Vert\bar\theta(t_{k})\Vert +T_{0}L_{2}\sqrt{V_{0}(t_{k})}\right)\notag\\
	&\leq L_{4}\sqrt{V_{0}(t_{k})}	  \label{eq-sqrtW1-case2}
\end{align}
where $L_{4}=\sqrt{\frac{\mu}{T\overline{C}_{0}}}\left((\frac{1}{T_{0}}+L_{1})\sqrt{\frac{2}{\lambda_{\min}(D)}} +T_{0}L_{2}\right)$ and $\Vert \bar \theta(t_{k})\Vert $ $\leq  \sqrt{\frac{2V_{0}(t_{k})}{\lambda_{\min}(D)}}$ are used in the last inequality, based on $V_{0}(t)$ in \eqref{V0}.
		
Now, we assume that there exists a subset  $\mathbb{S}$  of $\{t_k: k\geq k_0\}$ such that Case-2 holds at any sampling instant belongs to $\mathbb{S}$, while Case-1 holds at any other sampling instant belongs to $\{t_k: k\geq k_0\} / \mathbb{S}$. The following three situations are considered:
		
Situation-1: suppose that $\mathbb{S}$ is a finite set (or an empty set). In this situation,  let $t_{k^*}$ be the maximum number belongs to $\mathbb{S}$ (set $t_{k^*}=t_{k_0}$ if $\mathbb{S}$ is empty). Then, $\{t_k: k\geq k_0\} / \mathbb{S}$ is infinite, and Case-1 holds for any $k\geq k^*+1$. By \eqref{eq-chi}, $\sqrt{W_{1}(t_{k+1})}\leq \chi  \sqrt{W_{1}(t_{k})}$ holds for any $k\geq k^*+1$. It follows that $W_{1}(t_{k})$ converges to zero. By \eqref{sp-max2}, $W_{1M}(t)$, hence $W_{1}(t)$, converges to zero, and so do $ \bar x(t) $ and $ \bar y(t) $. 
		
Situation-2: suppose that  $\{t_k: k\geq k_0\} / \mathbb{S}$ is a finite set (or an empty set). In this situation,  let $t_{k^*}$ be the maximum number belongs to $\{t_k: k\geq k_0\} /\mathbb{S}$ (set $t_{k^*}=t_{k_0}$ if $\{t_k: k\geq k_0\} /\mathbb{S}$ is empty). Then, $\mathbb{S}$ is infinite, and Case-2 holds for any $k\geq k^*+1$. By \eqref{eq-sqrtW1-case2}, for any $t\in [t_{k}, t_{k+1})$ with $k \geq k^*+1$, 
$$
	\sqrt{W_{1}(t)} \leq L_{4}\sqrt{V_{0}(t_{k})} \leq   L_{4}  \varrho^{k-k^*-1} \sqrt{V_{0}(t_{k^*+1})}.
$$ 
Since $t \rightarrow \infty$ implies $k\rightarrow\infty$, we have  $W_{1}(t)$ converges to zero, so do $ \bar x(t) $ and $ \bar y(t) $.

Situation-3: suppose that $\mathbb{S}$ and $\{t_k: k\geq k_0\} / \mathbb{S}$ are both infinite sets. In this situation, Case-1 and Case-2 switch infinite times (at sampling instants $\{t_{k_0},t_{k_1},t_{k_2},\dots \}$) for all $t\geq t_{k_0}$. Without loss of generality, we assume that Case-2 holds  at $t_{k_{2l}}$ and Case-1 holds  at $t_{k_{2l+1}}$, where $l=0,1,2,\dots$. Then, for any $t\in [t_{k_{2l}}, t_{k_{2l+1}})$, by \eqref{sp-rho} and \eqref{eq-sqrtW1-case2},
\begin{align*}
	\sqrt{W_{1}(t)}
	&\leq L_{4}\sqrt{V_{0}(t_{k_{2l}}+\lfloor\frac{t - t_{k_{2l}}}{T_{0}}\rfloor T_{0})}
	\notag \\
	& 	\leq L_{4}\varrho^{{\lfloor\frac{t - t_{k_{2l}}}{T_{0}}\rfloor}}\sqrt{V_{0}(t_{k_{2l}})}<L_{4} \sqrt{V_{0}(t_{k_{2l}})}
\end{align*}
while, for any $t\in [t_{k_{2l+1}}, t_{k_{2l+2}})$, by \eqref{sp-rho}, \eqref{sp-max2}, \eqref{eq-chi}, and \eqref{eq-sqrtW1-case2},
\begin{align*}
	\sqrt{W_{1}(t)}
	&\leq \chi^{{\lfloor\frac{t - t_{k_{2l+1}}}{T_{0}}\rfloor}}\sqrt{W_{1}(t_{k_{2l+1}})}< \sqrt{W_{1}(t_{k_{2l+1}})} \\
	& \leq L_{4}\varrho^{\frac{t_{k_{2l+1}} - t_{k_{2l}}}{T_{0}}-1}\sqrt{V_{0}(t_{k_{2l}})} \leq L_{4}  \sqrt{V_{0}(t_{k_{2l}})}. 
\end{align*}
That is to say, for any $t\in [t_{k_{2l}}, t_{k_{2l+2}})$, by \eqref{sp-rho},
$
	\sqrt{W_{1}(t)} < L_{4} \sqrt{V_{0}(t_{k_{2l}})} \leq L_{4} \varrho^{k_{2l}-k_0} \sqrt{V_{0}(t_{k_0})} \leq L_{4} \varrho^{2l} \sqrt{V_{0}(t_{k_0})}.
$
Since $t \rightarrow \infty$ implies $l\rightarrow\infty$, we have $W_{1}(t)$ converges to zero, so do $ \bar x(t) $ and $ \bar y(t) $.

\textit{Step 3: Show that the origin of \eqref{SP-CL-00} is UGAS provided that $T_{0}$ satisfies $0<T_{0}< T^{*}$ with $T^{*}=\min\{T^{*}_1,T^{*}_2\}$.} From previous two steps, it remains to show that the origin of \eqref{SP-CL-00} is UGS. By \eqref{finite time-v1} and \eqref{ieq-w1} in Appendix, for any $t\in[0, t_{k_0})$,
\begin{align}\label{before-vv0}
	\sqrt{V_{1}(t)}&\leq \sqrt{W_{1}(t)}
	\leq \sqrt{\frac{\mu }{T\overline{C}_{0}}}\sqrt{V_{1}(t)}
	\notag\\
	&\leq\sqrt{\frac{\mu}{T\overline{C}_{0}}}\left((1+T_{0}L_{1})^{\lfloor\frac{t_{k_0} -t_{0}}{T_{0}}\rfloor+1}\sqrt{V_{1}(t_{0})}+\frac{L_{2}}{L_{1}} \right. \notag\\
	&~~~~~\left.  \times\left((1+T_{0}L_{1})^{\lfloor\frac{t_{k_0} -t_{0}}{T_{0}}\rfloor+1}-1\right) \sqrt{V_{0}(t_{0})}\right)\notag\\
	&\leq L_{5}\sqrt{V_{1}(t_{0})}+\frac{L_2L_5}{L_1}\sqrt{V_{0}(t_{0})}  
\end{align}
where $L_{5}=\sqrt{\frac{\mu}{T\overline{C}_{0}}}(1+T_{0}L_{1})^{\beta+1}$ with $\beta=\lfloor\frac{t_{k_0} -t_{0}}{T_{0}}\rfloor$. On the other hand, for $t\in[t_{k_0}, \infty)$, from the aforementioned three situations, 
\begin{align}\label{after-vv0}
	\sqrt{V_{1}(t)}\leq \sqrt{W_{1}(t)}\leq  L_{4}\sqrt{V_{0}(t_{k_0})}\leq L_{4}\sqrt{V_{0}(t_0)}
\end{align}
where $\max_{t\in[t_{0}, \infty)}\sqrt{V_{0}(t)}=\sqrt{V_{0}(t_{0})}$ is used in last inequality, based on the $V_{0}(t_{k+1})\leq \varrho^2 V_{0}(t_{k}) < V_{0}(t_{k}) $. Then, jointing \eqref{before-vv0} and \eqref{after-vv0} together gives that
\begin{align}\label{ugs-v1v0}
	\sqrt{V_{1}(t)}+\sqrt{V_{0}(t)}
	&\leq L_{5}\sqrt{V_{1}(t_{0})}+L_{6}\sqrt{V_{0}(t_{0})} 
\end{align}
where $L_{6}=\frac{L_2L_5}{L_1}+L_4+1$. Then, from \eqref{V0} and \eqref{SLF-V1}, 
\begin{align*}
	\sqrt{V_{1}(t)}+\sqrt{V_{0}(t)}
	&\geq \sqrt{\frac{\lambda_{\min}(D)}{2}}\Vert [\bar x^{\t}(t), \bar y^{\t}(t), \bar\theta^{\t}(t)]\Vert
\end{align*}
and
\begin{align*}
	&L_{5}\sqrt{V_{1}(t_{0})}+L_{6}\sqrt{V_{0}(t_{0})}	\\
	&\leq \sqrt{\frac{\lambda_{\max}(D)}{2}}(L_5\Vert[\bar x^{\t}(t_{0}), \bar y^{\t}(t_{0})]\Vert+L_6\Vert \bar\theta(t_{0}) \Vert)\notag\\
	&\leq \sqrt{\lambda_{\max}(D)}\max\{L_5, L_6\}\Vert [\bar x^{\t}(t_{0}), \bar y^{\t}(t_{0}), \bar\theta^{\t}(t_{0})]\Vert.
\end{align*}
Thus,
$
	\Vert [\bar x^{\t}(t), \bar y^{\t}(t), \bar\theta^{\t}(t)]\Vert
 \leq \sqrt{\frac{2\lambda_{\max}(D)}{\lambda_{\min}(D)}}\max\{L_{5}, L_{6}\}r_0
$,
where $r_0=\Vert [\bar x^{\t}(t_{0}), \bar y^{\t}(t_{0}), \bar\theta^{\t}(t_{0})]\Vert$. So  the origin of \eqref{SP-CL-00} is UGS, and hence UGAS.   
\end{Proof}

\begin{Remark}
Notice that the controller \eqref{cont-000} is piecewise continuous, relying on the exact time-varying signals $\omega_{0}(t)$ and $v_{0}(t)$, where the main purpose is to achieve exact asymptotic tracking for a time-varying leader's trajectory. Indeed, the sample-and-hold controller by direct making use of sampling values $\omega_{0}(t_k)$ and $v_{0}(t_k)$ cannot compensate the steady-state values exactly, hence undoubtedly suffers the so-called ripple phenomenon \cite{TAC-1986-Franklin-Naeini} and leads only to the  practical tracking, whenever $\omega_{0}(t)$ and $v_{0}(t)$ are not constant.  
\end{Remark}

\begin{Remark}
The proof of Theorem \ref{theorem-03} provides a novel sampled-data feedback control framework based on the strong iISS-Lyapunov function which aims to obtain the uniform global attractivity of $\bar x, \bar y$. In fact, inequality \eqref{sp-der-w1} means that system \eqref{SP-CL-02} is strongly iISS with respect to $\bar\theta(t_{k})$ and $\hat{x}(t)$. This property enables us to characterize the upper bound $T_2^*$ of the admissible sampling period by ensuring $W_{1}$ decreased from one sampling instance to the next, no matter what the sign of $\dot W_{1}$ in between sampling instants is, for any sufficiently large time $t$. Consequently, owing to the cascaded structure, the uniform global attractivity of $\bar x, \bar y$ can be achieved by evaluating the steady state of $\bar x, \bar y$ and $\bar \theta$ as in the three situations considered in the proof. 
\end{Remark}

\begin{Remark} 
Different from some previous results on the cooperative sampled-data feedback control, say  \cite{IJC-2019-PZX}, the communication network in this paper is allowed to be more general, i.e., directed and containing a spanning tree. In contrast to \cite{Auto-2017-LZX}, besides orientation and speed tracking, position tracking has further been achieved. 
\end{Remark}

\begin{figure}
	\centering
	\includegraphics[width=0.8\linewidth]{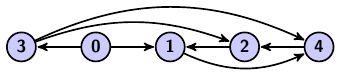}
	\caption{Communication topology graph $\BG$.}\label{graph}
	\vspace{-0.3cm}
\end{figure}
	
\begin{table}
	\caption{Initial values and desired position}\label{table}
	\vspace{-0.1cm}
	\centering
	\begin{tabular}{c|c|c|c|c|c}
		\hline
		\hline \diagbox{Agents}{Values}
		& $x_{i}(0)$ & $y_{i}(0)$  & $\theta_{i}(0)$ & $p^{x}_{i0}$ & $p^{y}_{i0}$ \\
		\hline
		0  &    0   & -6.25  & 0  &0 & 0\\
		\hline
		1  & -0.33  & -0.4   & -(1/6)$\pi$   & -1 & 0\\
		\hline
		2  & 0.92   & 0.2    &  (1/3)$\pi$   &  0 & 1\\
		\hline
		3  & -0.15  & 0.88   &  (2/3)$\pi$   &  1 &  0\\
		\hline
		4  & -1.18  & 1      &  (1/2)$\pi$   &  0 & -1\\ 
		\hline
	\end{tabular}
	\vspace{-0.3cm}
\end{table}

\section{Numerical example}\label{sec-example}
\begin{figure}
	\centering
	\scalebox{0.55}{\includegraphics[clip,bb=-10 0 405 320]{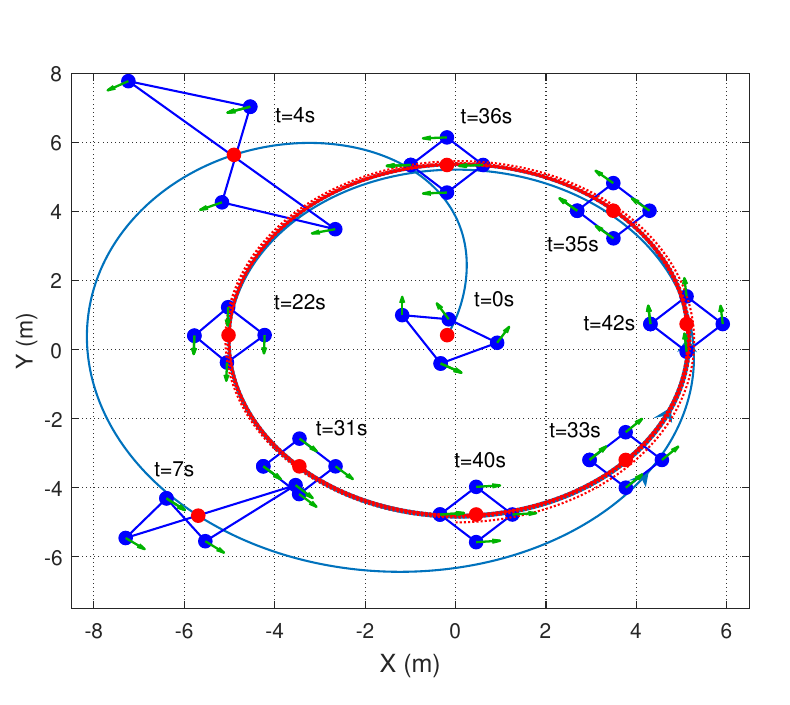}}\vspace{-0.3cm}
	\caption{The trajectory of the center of four follower robots (cyan solid line), the trajectory of leader robot (red dotted line), and their formation pattern.}\label{conref}
	\vspace{-0.3cm}
\end{figure}

\begin{figure}
	\centering
	\scalebox{0.57}{\includegraphics[clip,bb=-12 0 520 300]{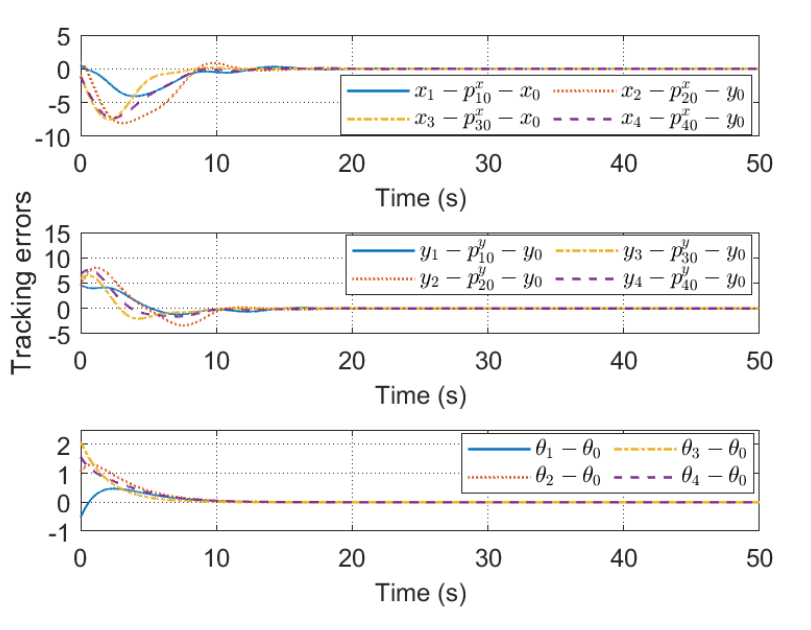}}\vspace{-0.3cm}
	\caption{The formation errors between each follower and leader.}\label{xyz}
	\vspace{-0.3cm}
\end{figure}

In this section, to illustrate the effectiveness of our designs, we present a numerical example of a group of four follower robots and one leader robot. The formation-tracking of multi-robot systems are considered, where relative positions between the followers and the leader are constants. The communication topology graph $\BG$ is shown in Fig.~\ref{graph}, satisfying Assumption \ref{ass-04}. The linear and angular velocities of the leader are set to  $\omega_{0}=0.8-\frac{1}{\sqrt{400t+800}}$ and $v_{0}=4+\frac{1}{\sqrt{100t+200}}$, satisfying Assumptions \ref{ass-02} and \ref{ass-03}. For simplicity, we only perform the distributed sampled-data controller \eqref{cont-000}. Set $k_{v}=1$, $k_{\omega}=0.5$, and  $T_{0}=0.04$. The initial positions and orientations for all multi robots and desired relative positions between the followers and the leader are given in Table \ref{table}. 
Fig.~\ref{conref} shows the evolution of the formation and the trajectory of the centroid of four follower systems. It can be seen that the desired geometric pattern is achieved and the centroid of follower systems converges to the trajectory of the leader. Fig.~\ref{xyz} depicts the formation tracking errors including position and orientation converge to zero. 
	
\begin{Remark}
According to  the proof of Theorem 3, as the selection of scaling and parameters may not be optimal, a sufficient but somewhat conservative design for $T^{*}$ is $T^{*}\triangleq\min\{T^{*}_1,T^{*}_2\}$, where
\begin{subequations}\label{T1T2}
		\begin{align*}
			T^{*}_{1}&=\frac{h_{1}}{h_{2}}=\frac{k_{\omega}\lambda_{\min}(Q)}{\lambda_{\max}(D)}\cdot \frac{\lambda_{\min}(D)}{2k^2_{\omega}\Vert H \Vert \Vert DH \Vert} \\
			T^*_2&=\min\left\{\frac{3\overline{C}_{0}}{4C_{2}}, \frac{-C_{2}-C_{4}L_{3}+\sqrt{(C_{2}+C_{4}L_{3})^2+3\overline{C}_{0}C_4C_{5} }}{2C_4C_{5}} \right\}.
		\end{align*}
\end{subequations}
Here, $k_{\omega}$, $D$, $H$, $Q$, $C_{2}$, $\overline{C}_{0}$ $C_{4}$, $L_{3}$ and $C_{5}$ are defined in \eqref{cont-00}, \eqref{der-V0}, \eqref{der-w1-02}, \eqref{eq-W-uuu}, \eqref{sp-der-w1}, and \eqref{add-boundx}, respectively. Indeed, in our example, with some calculations, it gives $T^{*}=9.1764\times 10^{-6}$, which is much more conservative than $T_{0}=0.04$ as setting here. 
\end{Remark}

\section{Conclusions}\label{sec-conclusion}
In this paper, the cooperative tracking problem of multiple nonholonomic mobile robots is investigated over general directed communication network. Firstly, uniform global asymptotic stability of the closed-loop system is achieved by the proposed distributed continuous control law,  where an explicit strict Lyapunov function is constructed. 
Secondly, the convergence rate of the closed-loop system is analyzed and it is proven that the closed-loop system is globally $\mathcal{K}$-exponentially stable. Thirdly, the distributed continuous control law is generalized to a distributed sampled-data control law by the emulation approach. Note that the communication network is assumed to be a static digraph in this paper. It is interesting to consider further the case of time-varying communication networks in the future. The cooperative tracking  using the event-triggered control strategy is also of great interest \cite{TAC-2023-ZPP,SCIS-Xu-Su-Cai-2023}, especially for the unbounded leader in the straight-line or even with non-PE trajectory.

\appendix

\begin{ProofL2}
Under Assumption \ref{ass-03} and the definition of $\varphi(t)$, $ 1 \leq \varphi(t)\leq 1+T\bar\omega$. Recalling Assumption \ref{ass-03} again, it follows that
$
\left|-\omega_{0}\bar x^{\t}D\bar y\right|\leq \frac{\bar \omega}{2}\bar x^{\t}D\bar x+\frac{\bar \omega}{2}\bar y^{\t}D\bar y
$.
This together with the fact that $\gamma\geq \bar\omega$ gives 
\begin{align}\label{ieq-w1}
	V_{1}\leq W_{1}(t,\bar x, \bar y)\leq \left(1+T\bar \omega+2\gamma\right)V_{1}.
\end{align}
So $W_{1}(t,\bar x,\bar y)$ is a positive definite, descent, and radially unbounded function.
		
Next, we consider the derivative of $W_{1}(t,\bar x,\bar y)$ along the trajectory of the nominal dynamics of \eqref{CL-02}. Simple calculations give that
\begin{align}\label{dW-eq}
	\dot{W}_1 \leq	
	& -\frac{k_{v}(1+\gamma)\lambda_{\min}(Q)}{2}\bar x^{\t}\bar x-\frac{2\mu}{T}V_{1} + 2 \omega_{0}^{2}(t)\bar x^{\t}D\bar x \nonumber\\
	& +k_{v}\omega_{0}(t)\bar x^{\t}H^{\t}D\bar y  -\dot \omega_{0}(t) \bar x^{\t}D\bar y.
\end{align}  
It can be verified that, for any $\epsilon>0$,
\begin{align*}
	\left|k_{v}\omega_{0}(t)\bar x^{\t}H^{\t}D\bar y\right|
	&\leq \frac{\epsilon k_{v}\bar \omega\Vert H \Vert^2}{2}\bar x^{\t}\bar x+\frac{k_{v}\bar \omega\lambda_{\max}(D)}{2\epsilon}\bar y^{\t}D\bar y\\
	\left|-\dot \omega_{0}(t) \bar x^{\t}D\bar y\right|
	&\leq \frac{\epsilon\bar \omega\lambda_{\max}(D)}{2}\bar x^{\t}\bar x+\frac{\bar \omega}{2\epsilon}\bar y^{\t}D\bar y.
\end{align*}
Then, with $\epsilon=\frac{T}{\mu}\left(\bar \omega+k_{v}\bar \omega\lambda_{\max}(D)\right)$, \eqref{dW-eq} implies that, along the trajectory of the nominal dynamics of \eqref{CL-02}, 
\begin{align}\label{der-w1-nominal}
	\dot W_{1}\leq -\frac{\mu}{T}V_{1}.
\end{align}
Now, we consider the derivative of $W_{1}(t,\bar x,\bar y)$ along the trajectory of \eqref{CL-02}. With \eqref{der-w1-nominal}, simple calculations give that 
\begin{align}\label{der-w1}
	\dot W_{1}
	& \leq -\frac{\mu}{T}V_{1}+(\phi(t)+\gamma)\frac{\partial V_{1}}{\partial [\bar x^{\t} \ \bar y^{\t}]^{\t}}\begin{bmatrix}	-k_{\omega}\mbox{diag}(\bar y)H\bar\theta \\ k_{\omega}\mbox{diag}(\bar x)H\bar\theta\end{bmatrix}\notag\\
	&~~~~+(\phi(t)+\gamma)\frac{\partial V_{1}}{\partial [\bar x^{\t} \ \bar y^{\t}]^{\t}}\begin{bmatrix}-k_{\omega}\tilde y_{0}H\bar\theta \\ k_{\omega}\tilde x_{0}H\bar\theta\end{bmatrix} \notag\\
	&~~~~-\omega_{0}(-k_{\omega}\mbox{diag}(\bar y)H\bar\theta)^{\t}D\bar y-\omega_{0}\bar x^{\t}D(k_{\omega}\mbox{diag}(\bar x)H\bar\theta)\notag\\
	&~~~~-\omega_{0}(-k_{\omega}\tilde y_{0}H\bar\theta)^{\t}D\bar y-\omega_{0}\bar x^{\t}D(k_{\omega}\tilde x_{0}H\bar\theta) .
\end{align}
Since $\mbox{diag}(\bar y)H\bar\theta=\mbox{diag}(H\bar\theta)\bar y$, it holds that	
\begin{align*}
	&\frac{\partial V_{1}}{\partial [\bar x^{\t} \ \bar y^{\t}]^{\t}}\begin{bmatrix}	-k_{\omega}\mbox{diag}(\bar y)H\bar\theta \\ k_{\omega}\mbox{diag}(\bar x)H\bar\theta\end{bmatrix}
	=\frac{\partial V_{1}}{\partial [\bar x^{\t} \ \bar y^{\t}]^{\t}}\begin{bmatrix}-k_{\omega}\mbox{diag}(H\bar\theta)\bar y \\ k_{\omega}\mbox{diag}(H\bar\theta)\bar x\end{bmatrix}\\
	&=[\bar x^{\t} \ \bar y^{\t}]\begin{bmatrix} D &  0 \\  0  & D  \end{bmatrix}
	\begin{bmatrix}0  &  -k_{\omega}\mbox{diag}(H\bar\theta) \\  k_{\omega}\mbox{diag}(H\bar\theta) & 0 \end{bmatrix}
	\begin{bmatrix} \bar x  \\  \bar y\end{bmatrix}=0.
\end{align*}
Then, 
\begin{align}
	\dot W_{1}
	&\leq 
	-\frac{\mu}{T}V_{1}-k_{\omega}\omega_{0}\left(\bar x^{\t}D\mbox{diag}(\bar x)-\bar y^{\t}D\mbox{diag}(\bar y)\right)H\bar\theta\notag\\
	&~~~~-k_{\omega}(\omega_{0}\tilde x_{0}+\left(\varphi(t)+\gamma)\tilde y_{0}\right)\bar x^{\t}DH\bar\theta\notag\\
	&~~~~+k_{\omega}(\omega_{0}\tilde y_{0}+(\varphi(t)+\gamma)\tilde x_{0})\bar y^{\t}DH\bar\theta\notag\\
	&\leq -\frac{\mu}{T}V_{1}+k_{\omega}\bar \omega\Vert H \Vert\Vert \bar x^{\t}D\mbox{diag}(\bar x)-\bar y^{\t}D\mbox{diag}(\bar y)\Vert\Vert \bar\theta \Vert +\sqrt{2}\notag\\
	&~~~\cdot k_{\omega}M(1+\gamma+T\bar\omega+\bar\omega)\Vert H \Vert(\Vert \bar x^{\t}D\Vert + \Vert \bar y^{\t}D\Vert)\Vert \bar\theta\Vert \label{eq-Waaa2}
\end{align}
where Assumptions \ref{ass-02}-\ref{ass-03} and the bound of $\varphi(t)$ are used in last inequality. Note that
\begin{align*}
	& 	\Vert \bar x^{\t}D\mbox{diag}(\bar x)-\bar y^{\t}D\mbox{diag}(\bar y)\Vert 
	\leq \Vert \bar x^{\t}D\mbox{diag}(\bar x)\Vert+\Vert\bar y^{\t}D\mbox{diag}(\bar y)\Vert\notag\\
    &~~~~~~~~	\leq \bar x^{\t}D\bar x+\bar y^{\t}D\bar y \leq  2V_{1}\\
	&\Vert x^{\t}D \Vert +\Vert y^{\t}D \Vert \leq \sqrt{\lambda_{\max}(D)}\left(\sqrt{\bar x^{\t}D\bar x} +\sqrt{\bar y^{\t}D\bar y}\right)\notag\\
	&~~~~~~~~\leq \sqrt{\lambda_{\max}(D)}\sqrt{2(\bar x^{\t}D\bar x+\bar y^{\t}D\bar y)} \leq 2\sqrt{\lambda_{\max}(D)V_{1}}.
\end{align*}
Letting $C_{1}=2k_{\omega}\bar\omega \Vert H \Vert$ and $C_{2}=2\sqrt{2}k_{\omega}M(1+\gamma+T\bar\omega+\bar\omega)\sqrt{\lambda_{\max}(D)}\Vert H \Vert$ in \eqref{eq-Waaa2} gives \eqref{der-w1-02}. 
\end{ProofL2}


\begin{thebibliography}{99}
\bibitem{book-2008-RW}
W. Ren and R. Beard, \textit{Distributed Consensus in Multi-Vehicle Cooperative Control}. London, U.K.: Springer-Verlag, 2008.
	
	
\bibitem{book-2009-QZH}
Z. Qu, \textit{Cooperative Control of Dynamical Systems: Applications to Autonomous Vehicles}. London: Springer-Verlag, 2009.
	
	
	
\bibitem{TAC-2008-DWJ}
W. Dong and J. A. Farrell, ``Cooperative control of multiple nonholonomic mobile agents,'' \textit{IEEE Trans. Autom. Control}, vol. 53, no. 6, pp. 1434--1448, Jul. 2008.
	
	
	
\bibitem{TR-2012-DWJ}
W. Dong, ``Tracking control of multiple-wheeled mobile robots with limited information of a desired trajectory,'' \textit{IEEE Trans. Robot.}, vol. 28, no. 1, pp. 262--268, Feb. 2012.	
	
	
	
\bibitem{IJSS-2015-PZX}
Z. Peng, G. Wen, A. Rahmani, and Y. Yu, ``Distributed consensus-based formation control for multiple nonholonomic mobile robots with a specified reference trajectory,'' \textit{Int. J. Syst. Sci.}, vol. 46, no. 8, pp. 1447--1457, 2015.
	
	

	

\bibitem{CSL-2022-WNA} 
B. Wang, S. Nersesov, and H. Ashrafiuon, ``Formation regulation and tracking control for nonholonomic mobile robot networks using polar coordinates,'' \textit{IEEE Control Syst. Lett.}, vol. 6, 1909--1914, Dec. 2022.




\bibitem{TAC-2023-YS} X. Yu and R. Su, ``Decentralized circular formation control of nonholonomic mobile robots under a directed sensor graph,'' \textit{IEEE Trans. Autom. Control}, vol. 68, no. 6, pp. 3656--3663, Jun. 2023.


	
\bibitem{Auto-2013-LTF}
T. Liu and Z.-P. Jiang, ``Distributed formation control of nonholonomic mobile robots without global position measurements,'' \textit{Automatica}, vol. 49, no. 2, pp. 592--600, 2013.
	
	
	
\bibitem{Auto-2014-LTF}
T. Liu and Z.-P. Jiang, ``Distributed nonlinear control of mobile autonomous multi-agents,'' \textit{Automatica}, vol. 50, no. 4, pp. 1075--1086, 2014.
	
	
	
\bibitem{TAC-2018-Loria}
M. Maghenem, A. Lor\'{\i}a, and E. Panteley, ``A cascades approach to formation-tracking stabilization of force-controlled autonomous vehicles,'' \textit{IEEE Trans. Autom. Control}, vol. 63, no. 8, pp. 2662--2669, Aug. 2018.
	
	
	
\bibitem{TAC-2020-Loria}
M. A. Maghenem, A. Lor\'{\i}a, and E. Panteley, ``Cascades-based leader–follower formation tracking and stabilization of multiple nonholonomic vehicles,'' \textit{IEEE Trans. Autom. Control}, vol. 65. no. 8, pp. 3639--3646, Aug. 2020.

	

\bibitem{ND-2017-TGP}
M. Tayefi, Z. Geng, and X. Peng, ``Coordinated tracking for multiple nonholonomic vehicles on SE (2),'' \textit{Nonlinear Dyn.}, vol. 87, pp. 665--675, 2017.

	
 
\bibitem{IJRNC-2021-HGeng}
X. He and Z. Geng, ``Trajectory tracking of nonholonomic mobile robots by geometric control on special Euclidean group,'' \textit{Int. J. Robust Nonlinear Control}, vol. 31, no. 12, pp. 5680--5707, 2021.



\bibitem{TAC-2018-NB}
B. Ning and Q.-L. Han, ``Prescribed finite-time consensus tracking for multiagent systems with nonholonomic chained-form dynamics,'' \textit{IEEE Trans. Autom. Control}, vol. 64, no. 4, pp. 1686--1693, Apr. 2019. 


 
\bibitem{TIE-2015-YL}
X. Yu and L. Liu, ``Distributed formation control of nonholonomic vehicles subject to velocity constraints,'' \textit{IEEE Trans. Ind. Electron.}, vol. 63. no. 2, pp. 1289--1298, Feb. 2016.
 
 
 
\bibitem{TASE-2018-MZQ}
Z. Miao, Y. H. Liu, Y. Wang, G. Yi, and R. Fierro, ``Distributed estimation and control for leader-following formations of nonholonomic mobile robots,'' \textit{IEEE Trans. Autom. Sci. Eng.}, vol. 15. no. 4, pp. 1946--1954, Oct. 2018.



\bibitem{IJC-2019-PZX}
X. Chu, Z. Peng, G. Wen, and A. Rahmani, ``Distributed formation tracking of nonholonomic autonomous vehicles via event-triggered and sampled-data method,'' \textit{Int. J. Control}, vol. 92, no. 10, pp. 2243--2254, 2019.
	
	
	
\bibitem{Auto-2017-LZX}
Z. Liu, L. Wang, J. Wang, D. Dong, and X. Hu, ``Distributed sampled-data control of nonholonomic multi-robot systems with proximity networks,'' \textit{Automatica}, vol. 77, pp. 170--179, 2017.
	
	
	
\bibitem{TAC-2023-ZPP}
P. Zhang, T. Liu, and Z.-P. Jiang, ``Tracking control of unicycle mobile robots with event-triggered and self-triggered feedback,'' \textit{IEEE Trans. Autom. Control}, vol. 68, no. 4, pp. 2261--2276, Apr. 2023.
	
	
	
\bibitem{SCIS-Xu-Su-Cai-2023}
L. Xu, Y. Su, and H. Cai, ``Event-triggered tracking control for a class of nonholonomic systems in chained form,'' \emph{Sci China Inf Sci}, vol. 66, no. 7, pp. 1--15, 2023.
	
	
	
\bibitem{book-2009-Mazenc}
M. Malisoff and F. Mazenc, \textit{Constructions of Strict Lyapunov Functions}. London: Springer-Verlag, 2009.
	
	
	
\bibitem{Auto-2017-Maghenem}
M. A. Maghenem and A. Lor\'{\i}a, ``Strict Lyapunov functions for time-varying systems with persistency of excitation,'' \textit{Automatica}, vol. 78, pp. 274--279, 2017.
	
	
	
\bibitem{TAC-2010-Ito}
H. Ito, ``A Lyapunov approach to cascade interconnection of integral input-to-state stable systems,'' \textit{IEEE Trans. Autom. Control}, vol. 55, no. 3, pp. 702--708, Mar. 2010.
	
	


\bibitem{CSL-2023-DLPSN}
M. Dutta, A. Lor\'{\i}a, E. Panteley, S. Sukumar, and E Nu\~{n}o, ``A $\delta$-persistently-exciting formation controller for non-holonomic systems over directed graphs,'' \textit{IEEE Control Syst. Lett.}, vol. 7, pp. 2587--2592, Jun. 2023.



\bibitem{TAC-2012-SYF}
Y. Su and J. Huang, ``Cooperative output regulation of linear multi-agent systems,'' \textit{IEEE Trans. Autom. Control}, vol. 57, no. 4, pp. 1062--1066, Apr. 2012.
	

	
\bibitem{TAC-2014-Ito}
A. Chaillet, D. Angeli, and H. Ito, ``Combining iISS and ISS with respect to small inputs: the strong iISS property,'' \textit{IEEE Trans. Autom. Control}, vol. 59. no. 9, pp. 2518--2524, Sep. 2014.
	
	
	
\bibitem{TAC-1986-Franklin-Naeini}
G. F. Franklin and A. Emami-Naeini, ``Design of ripple-free multivariable robust servomechanisms,'' \textit{IEEE Trans. Autom. Control}, vol. 31, no. 7, pp. 661--664, Jul. 1986.
	

	
\end{thebibliography}
\end{document}